\documentclass[aps,prd,
nofootinbib,
superscriptaddress,
tightenlines,
amsmath,
amssymb
]{revtex4}
\usepackage{bm}
\usepackage{amsmath}
\usepackage{amsfonts}
\usepackage{amssymb}
\usepackage{amssymb}
\usepackage{slashed}
\usepackage{color, graphicx}   
\usepackage{float}      
\usepackage{rotating}   
\usepackage{caption}    
\usepackage{subcaption} 
\usepackage{hyperref}

\usepackage{xcolor}
\usepackage{array}

\usepackage[normalem]{ulem}

\begin{document}

\title{Likelihood \& Correlation
Analysis of $\pi^o$ Electroproduction for Proton Tensor Charge Extraction }

\author{Saraswati Pandey}
\affiliation{Department of Physics, University of Virginia, Charlottesville, VA 22904, USA}

\author{Douglas Q. Adams}
\affiliation{Department of Physics, University of Virginia, Charlottesville, VA 22904, USA}

\author{Simonetta Liuti}
\affiliation{Department of Physics, University of Virginia, Charlottesville, VA 22904, USA}

\author{Zaki Panjsheeri}
\affiliation{Department of Physics, University of Virginia, Charlottesville, VA 22904, USA}

\author{Brannon Semp}
\affiliation{Department of Physics, University of Virginia, Charlottesville, VA 22904, USA}

\author{Kemal Tezgin}
\affiliation{Department of Physics, University of Virginia, Charlottesville, VA 22904, USA}

\begin{abstract}
We present a likelihood analysis of observables in deeply virtual exclusive $\pi^o$ electro-production off the proton. 
We use experimental data for the cross section as well as the beam and target spin asymmetries measured in various longitudinal polarization configurations at Jefferson Lab Hall A and Hall B. Performing a local extraction, for each kinematic bin, 
we derive a joint likelihood for the various helicity amplitudes defining the different polarization configurations. 
We subsequently extract the Compton form factors that parameterize the underlying QCD scattering amplitudes in the chiral-odd sector. 
Markov chain Monte Carlo (MCMC) methods are employed to explore the full parameter space, extracting the parameter uncertainties, correlations and credible intervals. 
The joint likelihood analysis of the cross-section and asymmetry data constrains the parameter space indicating that a larger kinematic coverage and additional polarization configurations, {\it e.g.} in the transverse sector, will improve the precision of the parameters.
We also report a determination of the proton tensor charge and tensor anomalous magnetic moment extracted in a Bayesian statistical framework using deeply virtual exclusive scattering data.
\end{abstract}

\date{April 2025}

\maketitle

\section{Introduction}

In exclusive deeply virtual meson production (DVMP, $ep \rightarrow e' p' M$) an electron scatters off a nucleon and a meson is detected in the final state along with the recoiling nucleon. In the one-photon exchange approximation, the momentum transfer to the nucleon target is given by the difference between the initial and final electron four-momenta, $q=k-k'$, with four-momemntum squared, $Q^2$ much larger than the proton mass squared, $M^2$ ($Q^2=-q^2>>M^2$). 
The scattering process is defined by two planes, chosen as the lepton plane where the initial and final electron momenta lie, and the hadron plane, where the final meson and nucleon emerge, forming an azimuthal angle $\phi$ with the lepton plane \cite{Raskin:1988kc,Diehl:2005pc}.
QCD factorization theorems are expected to hold in the limit where the momentum transfer squared between the initial and final nucleons, $t= \Delta^2 = (p'-p)^2$ is much smaller than $Q^2$ \cite{Collins:1996fb}, allowing the cross section to be written as the convolution of the meson distribution amplitude at the electro-production vertex, and the Compton form factors (CFFs) containing the nucleon generalized parton distributions (GPDs)  (we refer the reader to see Refs.\cite{Diehl:2003ny,Belitsky:2005qn,Kumericki:2016ehc} for reviews on the subject). 
\begin{figure}
    \centering
    \includegraphics[width=0.5\linewidth]{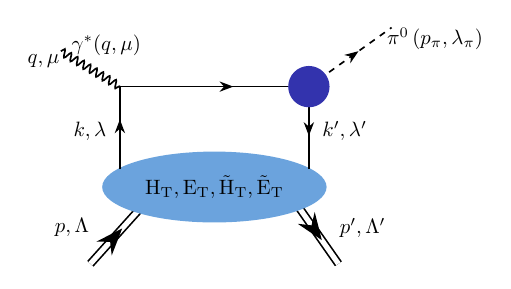}
    \caption{Leading order amplitude for the DVMP processes described in the text.}
    \label{fig:DVMP_process}
\end{figure}
As first shown in Ref.\cite{Ahmad:2008hp}, and subsequently developed and confirmed in Refs.\cite{Goldstein:2013gra,Goldstein:2012az,Goloskokov:2011rd,Goloskokov:2012bf,Goloskokov:2013mba,Duplancic:2016bge,Boussarie:2016aoq}, neutral pseudoscalar meson ($\pi, \eta, \eta',K, D$) electroproduction  Figure \ref{fig:DVMP_process}, provides a direct probe of the chiral-odd sector of nucleon structure, through which the otherwise elusive transversity/chiral-odd GPDs can be constrained. The latter were defined as, $H_T$, $E_T$, $\widetilde H_T$, and $\widetilde E_T$, thus encoding all allowed quark helicity-flip transitions at leading twist \cite{Diehl:2001pm}.

The work in Ref.\cite{Ahmad:2008hp} reformulated the earlier description of neutral pion electroproduction, which was solely based on the chiral-even GPDs,  $\widetilde{H}$ and $\widetilde{E}$ \cite{Mankiewicz:1998kg,Mankiewicz:1997uy,Vanderhaeghen:1998uc,Vanderhaeghen:1999xj},  
 by considering the $J^{PC}$ quantum numbers of the dominant $t$-channel exchanges in the $\gamma^* p \rightarrow \pi^0 p'$ process.
In particular, the $J^{PC}=1^{--}$ and $1^{+-}$ exchanges favor the ($\gamma_5$) $\pi^o$ coupling, or the chiral-odd mechanism, while the the conventional $\gamma_\mu (\gamma_5$) chiral-even mechanism, with $J^{PC}=1^{++}$, is suppressed.
In this respect, the working hypothesis introduced in Refs.~\cite{Ahmad:2008hp,Goldstein:2013gra} can be viewed as going beyond a simple collinear-factorization picture, through a two-step mechanism involving a diffractive transition followed by a collinearly factorized hard subprocess. Within such a picture, angular correlations between the hadronic diffractive plane and the hard-scattering plane provide sensitivity to the spin structure of the exchanged state. A broader formulation of this type of framework, encompassing a wider class of final states and crossed channels, has more recently been proposed in Ref.~\cite{Qiu:2022pla}.
%
Besides its theoretical underpinnings, the chiral-odd mechanism successfully provided an initial explanation for the large transverse photon polarization contributions to the cross section ($\sigma_T$, $\sigma_{TT}$, $\sigma_{LT}$) measured at Jefferson Lab over a broad kinematic range \cite{Bedlinskiy:2012be}.
Nevertheless, 
the interpretation of DV$\pi^0$P, remains theoretically an open question, since the observed dominance of transverse-photon amplitudes 
lies outside the leading-twist longitudinal regime in which standard collinear factorization for exclusive meson production has been established.
With this work, we revisit this question providing a statistical framework where both the role of chiral-odd GPDs and of factorization in the experimentally accessible kinematic region can be further investigated.

While chiral-odd GPDs provide unique information on the transverse-spin and spin-orbit QCD correlations in the proton,  an additional value of this analysis is that it provides an alternative route to extracting the tensor charge, rather than obtaining it through the underlying $x$-dependent transversity distribution, $h_1(x,Q^2)$ \cite{Ye:2016prn}, review in Ref.\cite{Barone:2001sp}.
Knowing the hadronic tensor charge is needed, in particular, in searches for Beyond-the-Standard-Model (BSM) scalar and tensor interactions in neutron $\beta$ decay \cite{Beck:2023hnt}, and high-energy processes at the LHC . In the conventional approach, the hadronic tensor charge is obtained as the first moment of the transversity distribution,
\begin{equation}
    \int_{-1}^{1} dx\, H_T(x,0,0)  = g_T 
    \label{eq:def_tensor_charge}
\end{equation}
requiring knowledge of the full $x$-dependence. By contrast, in deeply virtual exclusive processes the observables depend on chiral-odd Compton form factors (CFFs), which are already convolution integrals over $x$. This opens the possibility of constraining the tensor charge directly through CFF extractions, without first reconstructing the complete $x$-dependent GPD.
A precise knowledge of the mechanisms describing deeply virtual pseudoscalar meson electroproduction can impact also the outcome of recent studies such as the one conducted in Ref.\cite{Balkin:2026whv} focused on the search for axionlike particles and vector mediators/dark photons using existing photo and electroproduction data. 

The exclusive pseudoscalar channel in the deep-inelastic regime (with final state invariant mass, $W \gtrsim 2$ GeV) was studied at Jefferson Lab with CLAS using a longitudinally polarized $5.9$ GeV electron beam and a longitudinally polarized proton target. In Ref.~\cite{Kim:2015pkf}, target and double-spin asymmetries for deeply virtual exclusive $\pi^0$ electroproduction were measured over a broad kinematic range. 
More recently, CLAS12 measured the beam-spin asymmetry for deeply virtual exclusive $\pi^0$ production over a wide kinematic region, extending to photon virtualities of $Q^2=6.6~\mathrm{GeV}^2$ and covering $x_{Bj}$ in the valence regime \cite{CLAS:2023wda}. These measurements were performed with a longitudinally polarized $10.6$ GeV electron beam incident on an unpolarized liquid-hydrogen target.
Hall A measurements of the exclusive neutral-pion electroproduction cross section off the proton at large values of $x_{Bj}$ and $Q^2$ were presented in Ref.\cite{JeffersonLabHallA:2020dhq}. The measurements covered $x_{Bj}=0.36$, $0.48$, and $0.60$ and $Q^2$ from $3.1$ to $8.4~\mathrm{GeV}^2$. 
The results showed that amplitudes associated with transversely polarized virtual photons continue to dominate the cross section throughout the measured kinematic range. The data were also found to be well described by calculations based on transversity GPDs coupled to a helicity-flip pion distribution amplitude, supporting the use of exclusive $\pi^0$ electroproduction as a probe of the chiral-odd sector.

To quantify the information that can be extracted from the presently available data, we present a Bayesian likelihood and correlation analysis to extract the chiral-odd CFFs from deeply virtual exclusive pseudoscalar meson production. Bayesian inference methods are well established in fields such as astronomy and astrophysics and have increasingly been adopted in nuclear and hadronic physics, where they provide a systematic framework for parameter estimation, uncertainty quantification, and the determination of correlations among extracted quantities. Recently, this methodology was applied to the extraction of chiral-even CFFs from deeply virtual Compton scattering (DVCS) data, yielding well-constrained posterior distributions and revealing nontrivial correlations among the CFFs \cite{Adams:2024pxw}. Here, we extend this framework to the considerably less explored chiral-odd sector. However, in contrast to our previous likelihood analysis, which used the measured observables point by point in $\phi$, the present analysis uses the $\phi$-independent quantities extracted from their azimuthal dependence.
Consequently, the dataset entering the present analysis is less abundant and contains less direct information than the original $\phi$-dependent measurements. Ideally, the likelihood analysis would be performed directly on the measured $\phi$-dependent observables, allowing the azimuthal dependence and its correlations to be treated consistently within the inference procedure rather than relying on quantities obtained through an intermediate experimental extraction.

The paper is organized as follows: in Section \ref{sec:framework} we give a theoretical description of the cross section using a QCD-based formulation. We provide in-depth detail of the Bayesian method employed for our analysis along with the experimental data available for our analysis in Section \ref{sec:MCMC}; in Section \ref{sec:results}, we present our results including a first quantitative extraction of the chiral-odd CFFs.
We also examine the extent to which the available Hall A and Hall B data from Jefferson Lab support a description in terms of chiral-odd GPDs, and determine the constraints that these measurements place on the individual contributions in the chiral-odd sector.
Finally, we summarize our main results and give our conclusions and an outlook on future work Sec.~\ref{sec:summary}.

\section{Theoretical Framework and Motivation}
\label{sec:framework}
We consider the process $e p \rightarrow e'p' M$ where $M=\pi^o, \eta, ... $ is a neutral pseudoscalar meson (Fig.\ref{fig:DVMP_process}). 
We start by defining GPDs at twist-two as the matrix elements of  the following projection of the unintegrated quark-quark proton correlator (see Ref.\cite{Meissner:2007rx} for a detailed overview),
\footnote{In what follows we can omit the Wilson gauge link without loss of generality.}  
\begin{eqnarray}
W_{\Lambda', \Lambda}^\Gamma(x,\xi,t) & = & \frac{1}{2} \int \frac{d z^- }{2 \pi} e^{ix P^+ z^-} \left. \langle p', \Lambda' \mid \overline{\psi}\left(-\frac{z}{2}\right) \Gamma \, \psi\left(\frac{z}{2}\right)\mid p, \Lambda \rangle \right|_{z^+=0,{\bf z}_T=0},
\label{matrix}
\end{eqnarray}
where $\Gamma=\gamma^+, \gamma^+\gamma_5, i\sigma^{i+}\gamma_5 (i=1,2)$, and the target's spins are $\Lambda, \Lambda^\prime$. For the chiral-odd case, $\Gamma= i\sigma^{i+}\gamma_5$, $W_{\Lambda', \Lambda}^\Gamma$ can be parametrized as \cite{Diehl:2001pm},
\begin{eqnarray}
\label{correlator}
W_{\Lambda', \Lambda}^{[i\sigma^{i+}\gamma_5]}(x,\xi,t)  & = &  \frac{1}{2 {P}^+} \overline{U}(p',\Lambda') \left( i \sigma^{+i} H_T(x,\xi,t) +
 \frac{\gamma^+ \Delta^i - \Delta^+ \gamma^i}{2M} E_T(x,\xi,t)   \right. \nonumber \\
& + & \left.  \frac{p^+ \Delta^i - \Delta^+ p^i}{M^2}  \widetilde{H}_T(x,\xi,t)  +
\frac{\gamma^+ p^i - p^+ \gamma^i}{2M} \widetilde{E}_T(x,\xi,t) \right) U(p,\Lambda)
\end{eqnarray}
GPDs depend on four kinematic variables namely: $x$, and $t$, where $x$ is the average parton longitudinal momentum fraction and $\xi$ (skewness) representing half of the longitudinal momentum fraction transferred to the struck parton, is defined in terms of the Bjorken variable $x_B$ as $\xi = x_B/(2 - x_B)$, in which $x_B = Q^2 / 2p.q$, q is the four-momentum of the virtual photon, and $Q^2 = -q^2$. The momentum transfer to the nucleon is $t = (p - p')^2$, where p and $p'$ are the initial and final four momenta of the nucleon. 
The cross section is obtained by contracting the hadronic and leptonic tensor.
As shown in detail in Ref.\cite{Goldstein:2013gra}, it can be written in terms of structure functions as follows,
\begin{eqnarray} 
\label{xs}
\frac{d \sigma}{dx_{Bj} dy d\phi dt} & = & \Gamma \left\{ \left[ F_{UU,T} + \epsilon F_{UU,L}+ \epsilon \cos 2\phi F_{UU}^{\cos 2 \phi} 
+ \sqrt{\epsilon(\epsilon+1)} \cos \phi F_{UU}^{\cos \phi}   + 
h  \, \sqrt{\epsilon(1-\epsilon)} \, \sin \phi F_{LU}^{\sin \phi} \right] \right. \nonumber \\
& + & S_{||} \left[   \sqrt{\epsilon(\epsilon+1)} \sin \phi F_{UL}^{\sin \phi}  + \epsilon \sin 2 \phi F_{UL}^{\sin 2 \phi}  + h \, 
\left( \sqrt{1 - \epsilon^2} \, F_{LL} +    \sqrt{\epsilon(1-\epsilon)} \, \cos \phi \, F_{LL}^{\cos \phi}  \right) \right]   \nonumber \\
%
& - &  S_\perp \left[ \sin(\phi-\phi_S) \left(F_{UT,T}^{\sin(\phi-\phi_S)} + \epsilon  F_{UT,L}^{\sin(\phi-\phi_S)}  \right) + 
\frac{\epsilon}{2} \left( \sin(\phi+\phi_S) F_{UT}^{\sin(\phi+\phi_S)}  +  \sin(3\phi-\phi_S) F_{UT}^{\sin(3\phi-\phi_S)} \right)  \right. \nonumber \\
& + & \left. \sqrt{\epsilon(1+\epsilon)} \left( \sin\phi_S F_{UT}^{\sin \phi_S} + \sin(2\phi-\phi_S) F_{UT}^{\sin(2\phi-\phi_S)} \right) \right]   \nonumber \\
& + & \left. S_\perp h \left[ \sqrt{1-\epsilon^2} \cos(\phi-\phi_S) F_{LT}^{\cos(\phi-\phi_S)} +
\sqrt{\epsilon(1-\epsilon)} \left(\cos \phi_S F_{LT}^{\cos\phi_S} + \cos(2\phi-\phi_S) F_{LT}^{ \cos(2\phi-\phi_S)} \right) \right] \right\} \nonumber \\
\end{eqnarray}
where 
$S_{||}$ and ${\bf S}_\perp$ refer to lab frame target polarization parallel and perpendicular to the virtual photon direction, $h$ is the lepton beam helicity, $\phi$ is the azimuthal angle between the lepton plane and the hadron scattering plane, $\phi_S$ is the azimuthal angle of the transverse spin vector ${\bf S}_\perp$ and $t$ is the square of the invariant momentum transfer between the initial and final nucleon.
%

The photon polarization parameter $\epsilon$, the ratio of longitudinal photon and transverse photon flux, can be written in terms of invariants as \cite{Kriesten:2019jep},
\begin{equation}
\epsilon\equiv \frac{1-y-\frac{1}{4}y^2\gamma^2}{1-y+\frac{1}{2}y^2 +\frac{1}{4}y^2\gamma^2}
\label{eq:epsilon}
\end{equation}
\noindent 
$\Gamma$ is given by,
\begin{equation}
\label{Gamma}
\Gamma = \frac{\alpha^2 \, y^2 (1-x_{Bj})}{2\pi x_{Bj}(1-\epsilon)Q^2}.
\end{equation}
%
%
%
%
The kinematic variables which are essential for writing the cross section and various spin asymmetries are,
\begin{equation}
    [E, x_{Bj}, Q^2, t] 
    \label{eq:kinematics}
\end{equation}
where $E$ is the electron beam energy, $Q^2$ is the four-momentum transfer squared, $x_{Bj}=Q^2/2M\nu$, with $\nu$ the energy transfer, $t$ is the four momentum transfer squared between the initial and final protons.
Alternative notations that are frequently used in the literature are,
\begin{eqnarray}
F_{UU,T}  =  \frac{d\sigma_T}{dt} = \sigma_T,  \;\; F_{UU,L}=\frac{d\sigma_L}{dt} = \sigma_L, \;\;     F_{UU}^{\cos \phi} =\frac{d\sigma_{LT}}{dt} = \sigma_{LT},  
\;\;F_{UU}^{\cos 2\phi}  =   \frac{d\sigma_{TT}}{dt} = \sigma_{TT},  \;\; F_{LU}^{\sin \phi}  =   \frac{d\sigma_{LT'}}{dt} = \sigma_{LT'},  
\label{xsecs}
\end{eqnarray}
The various structure functions in Eq.(\ref{xs}) are written in terms of helicity amplitudes, $f_{\Lambda \Lambda'}^{\Lambda_\gamma^* \Lambda_{\gamma'}}$ \cite{Goldstein:2013gra,Goldstein:2012az}. The latter satisfy the following parity relations \cite{Jacob:1959at,Leader},
\begin{eqnarray}
f_{-\Lambda \gamma 0}^{-\Lambda -\Lambda'} & = & [\eta_G (-1)^{(\Lambda_{\gamma} - \Lambda)+\Lambda'} ] f_{\Lambda_\gamma 0}^{\Lambda \Lambda'}, \\
 \eta_G & = & (\eta_{\pi^o} \eta_{p'})/(\eta_\gamma \eta_p) (-1)^{S_{\pi^o}+ S_{p'} - S_\gamma - S_p}  = -1   
\end{eqnarray}
where $\eta_G$ is the intrinsic parities factor (see Eq.(43) in Ref.\cite{Jacob:1959at}). Therefore, one has only four helicity amplitudes for transverse and two for longitudinal virtual photon polarization, resulting in the following structure functions structures, 

The amplitudes composition in terms of helicity amplitude are given in the formulae below for the unpolarized nucleon target, as well as in the longitudinally and transversely polarized case. Note that our analysis will be, however, limited to using observables for the unpolarized and longitudinally polarized measurements, since data with a transversely polarized target are not yet available.  
\\

\noindent {\it Unpolarized target}
\begin{subequations}
\begin{eqnarray}
\label{dsigT}
F_{UU,T} & = &  \frac{1}{2} \left( \mid f^{10}_{++} \mid^2 + \mid f^{10}_{+-}  \mid^2 + \mid f^{10}_{-+}  \mid^2 +
\mid f^{10}_{--}  \mid^2 \right) 
\\ 
\nonumber \\
\label{dsigL}
F_{UU,L} & = &   \mid f^{00}_{+-}  \mid^2 + \mid f^{00}_{++}  \mid^2  \\
\nonumber \\
\label{dsigTT}
F_{UU}^{\cos 2 \phi} & = &  -  \,    \Re e  \left[  (f^{10}_{++})^ * (f^{10}_{--}) - (f^{10}_{+-})^* (f^{10}_{-+}) \right] \\
\nonumber \\
\label{dsigLT}
F_{UU}^{\cos \phi}  & = &  \Re e \! \left[   (f^{00}_{+-})^*(f^{10}_{+-} + f^{10}_{-+}) + (f^{00}_{++})^* (f^{10}_{++} - f^{10}_{--}) \right] \\
\nonumber \\
F_{LU}^{\sin \phi} & = & - \Im m  \! \left[  (f^{00}_{+-})^*(f^{10}_{+-} + f^{10}_{-+}) +  (f^{00}_{++})^* (f^{10}_{++} - f^{10}_{--}) \right]  
\label{ULsinphi:eq}
\end{eqnarray}
\end{subequations}
\\

\noindent{\it Longitudinally polarized target}
\begin{subequations}
\begin{eqnarray}
F_{UL}^{\sin \phi} & = &  
- \Im m \! \left[   (f^{00}_{+-})^*(f^{10}_{+-} - f^{10}_{-+}) + (f^{00}_{++})^* (f^{10}_{++} + f^{10}_{--})  \right] \\
\nonumber \\
\label{ULsin2phi:eq}
F_{UL}^{\sin 2 \phi} & = & -     \Im m \left[  (f^{10}_{++})^ * (f^{10}_{--}) - (f^{10}_{+-})^* (f^{10}_{-+}) \right] \\
\nonumber \\
\label{LLcosphi:eq}
F_{LL}^{\cos \phi} & = &   \Re e \! \left[   (f^{00}_{+-})^*(f^{10}_{+-} - f^{10}_{-+}) +  (f^{00}_{++})^* (f^{10}_{++} + f^{10}_{--})  \right] \\
\nonumber \\
\label{LL:eq}
F_{LL} & = &  \frac{1}{2} \left[  \mid f^{10}_{++} \mid^2 + \mid f^{10}_{+-} \mid^2 - \mid f^{10}_{-+} \mid^2 - \mid f^{10}_{--} \mid^2 
\right] 
\label{UL_LL}
\end{eqnarray}
\end{subequations}
\\

\noindent {\it Transversely polarized target}

\begin{subequations}
\begin{eqnarray}
F_{UT,T}^{\sin(\phi-\phi_S)} &=&   \Im m \left[ f_{10}^{++*} f_{10}^{-+} + f_{10}^{+-*} f_{10}^{--} \right] \\
F_{UT,L}^{\sin(\phi-\phi_S)} &=&  
 \, \Im m \left[ f_{00}^{++*} f_{00}^{-+} + f_{00}^{+-*} f_{00}^{--} \right]   \\
F_{UT}^{\sin(\phi+\phi_S)} &=& 
 \, \Im m \left[ -f_{10}^{++*} f_{10}^{+-} + f_{10}^{+-*} f_{10}^{++}  \right]  \\
F_{UT}^{\sin(3\phi+\phi_S)} &=& 
  \, \Im m \left[ f_{10}^{-+*} f_{10}^{--} - f_{10}^{--*} f_{10}^{-+}  \right] \\
F_{UT}^{\sin\phi_S} &=& 
   \, \Im m \left[ f_{10}^{++*} f_{00}^{-+} + f_{10}^{+-*} f_{00}^{--} \right] \\
F_{UT}^{\sin(2\phi-\phi_S)} &=&  
 \, \Im m \left[ f_{10}^{-+*} f_{00}^{++} + f_{10}^{--*} f_{00}^{+-} \right], \\
\label{UT}
F_{LT}^{\cos(\phi-\phi_S)} &=&    \, \Re e \left[ f_{10}^{++*} f_{10}^{-+} + f_{10}^{+-*} f_{10}^{--} \right]
\label{LT} \\
F_{LT}^{\cos\phi_S} &=&  
 \, \Re e \left[ f_{10}^{++*} f_{00}^{+-} + f_{10}^{+-*} f_{00}^{--} \right] \\
F_{LT}^{\cos(2\phi-\phi_S)} &=&    
 \, \Re e \left[ f_{10}^{-+*} f_{00}^{++} + f_{10}^{--*} f_{00}^{+-} \right] .
\end{eqnarray}
\end{subequations}
Knowledge of these quantities would highly benefit the amplitudes and CFFs extraction.
\\

\noindent The leading order helicity amplitudes for transverse photon polarization, are written in terms of chiral Odd Compton form factors (CFFs) as, (we denote the helicity amplitudes  by: $f_i$, $i=1,4$ for simplicity) \cite{Diehl:2001pm,Ahmad:2008hp,Goldstein:2013gra},
\begin{subequations}
\label{helamps_gpd2}
\begin{eqnarray}
\tilde{f}^{10}_{++} \equiv \tilde{f_1} &= &  g_1(Q^2) \,  e^{ i\phi}  \, \frac{\sqrt{t_0-t}}{4M} \left[ 2\tilde{ \cal H}_T  + (1+\xi)  \left({\cal  E}_T - \tilde{\cal  E}_T \right) \right] \\
\\
\tilde{f}^{10}_{+-} \equiv \tilde{f}_2 & = &  g_2(Q^2) \, \sqrt{1-\xi^2}  \left[ { \cal H}_ T + \frac{t_0-t}{4M^2} \tilde{ \cal H}_T      
- \frac{\xi^2}{1-\xi^2}  {\cal  E}_T  + \frac{\xi}{1-\xi^2} \widetilde{\cal  E}_T \right]  \nonumber \\ 
\tilde{f}^{10}_{-+} \equiv \tilde{f}_3 & = &  g_3(Q^2)   \, e^{- i 2 \phi}  \, \sqrt{1-\xi^2}  \,  \frac{t_0-t}{4M^2} \, \tilde{\cal  H}_T \nonumber \\
\tilde{f}^{10}_{--} \equiv \tilde{f}_4 & = &  g_4(Q^2) \,  e^{ i\phi}  \, \frac{\sqrt{t_0-t}}{4M}  \,  \left[  2\tilde{\cal H}_ T + (1-\xi) \left( { \cal E}_T + \tilde{\cal E}_T \right) \right]
\end{eqnarray} 
\end{subequations}
where the dependence on the phase angle, $\phi$,  of the various helicity amplitudes contributing to the cross section \cite{Jacob:1959at,Leader} given by,
\begin{equation}
\tilde{f}_{\Lambda \Lambda^\prime}^{\Lambda_{\gamma^*} \Lambda_\gamma'} (\theta,\phi) = e^{i(\Lambda_{\gamma^*} -\Lambda -\Lambda_\gamma'+ \Lambda')\phi} \,{f}_{\Lambda \Lambda^\prime}^{\,\Lambda_{\gamma^*} \Lambda_\gamma'}(\theta).
\end{equation}
was factored in the structure functions expressions, yielding the characteristic dependence of the cross section in Eq.\eqref{xs} (see Ref.\cite{Kriesten:2019jep} for a more detailed discussion). 
%
The CFFs are convolutions of the GPDs with pQCD kernels $C^{\pm}$ given by \cite{Diehl:2001pm}, 
\begin{equation}
\label{CFF_def}
    \widetilde{{\cal F}}(\xi,t,Q^2) = \int_{-1}^1 dx \; C^+(x,\xi) \, F_T(x,\xi ,t,Q^2) .
\end{equation}

Similarly to the chiral-even sector, all GPDs depend on the invariants $(x,Q^2,\xi,t)$, where $x$ is the momentum fraction carried by the partons, $Q^2$ is the virtual photon four-momentum squared defining the large scale for the deeply virtual scattering process, $\xi$ and $t$ described the longitudinal (along the light-cone) and transverse momentum transfers between the in initial and final proton.  

The four twist-two chiral-odd GPDs carry different information on the transverse spin structure of the nucleon: $H_T$ is the only chiral-odd GPD admitting a physically meaningful forward limit ($\xi,t \rightarrow 0$), given by the transversity distribution $h_1(x,Q^2)$. Furthermore, the first moment of $H_T$, evaluated at $t=0$, is identified with the hadronic tensor charge. 
As first shown in \cite{Burkardt:2005hp}, the first moment of the combination, $2 \tilde{H}_T + E_T$ defines the ``transverse anomalous magnetic moment"  of the proton. This combination therefore takes on a similar role as the GPD $E$ in the chiral-even sector, and it is responsible for dipole type deformations of the spatial density distribution.    
$\tilde{H}_T$ contributes to a quadrupole distortion of the transverse-spin density \cite{Diehl:2005jf}: by itself, being described by a kinematically suppressed double helicity flip distribution, it is difficult to isolate experimentally, but in impact-parameter space it governs anisotropic deformations of the quark distribution that depend on both the quark and proton transverse spins.
$\tilde{E}_T$ is qualitatively different: since it is odd in $\xi$, it disappears at zero skewness and therefore has no probability-density interpretation in impact-parameter space. Instead, it encodes genuinely off-forward spin-orbit correlations that become relevant only when longitudinal momentum is transferred.

The pion vertex functions are denoted by $g_i(Q^2)$, $i=1,4$,
\begin{equation}
 g_1=    g_\pi^{V, odd}(Q) , \quad g_2 =   \frac{g_\pi^{V, odd}(Q)+ g_\pi^{A, odd}(Q)}{2} , \quad g_3 =  \frac{g_\pi^{V, odd}(Q)- g_\pi^{A, Odd}(Q)}{2}  ,  \quad g_4 =  g_\pi^{V, odd}(Q) .
\end{equation}
The two distinct contributions in the vector ($V$) and axial-vector ($A$) sectors arise owing to the fact that, as observed in Ref.~\cite{Ahmad:2008hp},  the chiral-odd coupling $\propto \gamma^5$  contributes to $\pi^o$ electro-production provided one goes beyond a simple one gluon exchange description of this vertex.    
In the $t$-channel picture, which has its roots in a Regge analysis of this process \cite{Goldstein:1973xn}, one separates 
the $J^{PC}=1^{- -}$ and $J^{PC}=1^{+ -}$
contributions to the amplitudes for transverse and longitudinal virtual photons, respectively, thus generating two different types of $Q^2$ dependence at the $\pi^o$ 
vertex. The amplitudes in the chiral-even sector dominate the longitudinal photon polarization components of the cross section and they are given below. The functions $g_\pi^{V, odd}(Q)$, and  $g_\pi^{A, odd}(Q)$ were calculated in Refs.\cite{Ahmad:2008hp,Goldstein:2012az}. Their functional forms are given in Appendix \ref{app_g} for completeness. 
\\

\noindent {\it chiral-even Sector}
\\

\noindent 
We assume the $\pi^o$ electro-production amplitudes for longitudinal photon polarization to be dominated by the  twist-two chiral even GPDs in the axialvector sector \cite{Goeke:2001tz} (following the notation of the transverse case, here we introduce the notation: $f_5 = f^{00}_{+-}$, $f_6 = f^{00}_{++}$), 
\begin{subequations}
\label{f_even_1}
\begin{eqnarray}
f_{00}^{++, \, even} & =  &  f_5 = (1+\xi)\sqrt{\frac{1-\xi}{1+\xi}} \Bigg[\:\widetilde{\cal H}  -  \Big(\frac{\xi^2}{1-\xi^2}\Big)\:\widetilde{\cal E}\Bigg] , \\
f_{00}^{+-, \, even} & = &  f_6 = \xi \frac{\sqrt{t_o-t}}{2M} \: \widetilde{\cal E}, 
\end{eqnarray}
\end{subequations}

where $\widetilde{{\cal H}}, \, \, \widetilde{{\cal E}}$ are the CFFs corresponding to the GPDs $\tilde{H}$ and $\tilde{E}$ defined as, 
\begin{equation} \label{CFF_defi}
    \widetilde{\cal F} (\zeta,t,Q^2) = \int_{-1+\zeta}^1 dX \; C^+ \, F(X,\zeta ,t,Q^2) .
\end{equation}
\\

The measured quantities used in our analysis are the cross section terms and asymmetries written below
\begin{subequations}
\label{eq:allosbervables}
\begin{eqnarray}
 \frac{1}{\Gamma }  \sigma_{UU} & = &   \big({ F_{UU,T}} + \epsilon \, { F_{UU,L}} \big) + \epsilon \, \cos 2\phi \, { F_{UU}^{\cos 2 \phi}}
+ \sqrt{\epsilon(\epsilon+1)} \cos \phi \, { F_{UU}^{\cos \phi} }  + 
h  \, \sqrt{\epsilon(1-\epsilon)} \, \sin \phi \, { F_{LU}^{\sin \phi} }  \\
\nonumber \\
\label{ALU}
{ A_{LU}} , &&  \quad\quad A_{UL}  =  { A_{UL}^{\sin \phi} }  \sin \phi  + { A_{UL}^{\sin 2 \phi} } \sin 2 \phi,    \qquad\quad A_{LL}  =   { A_{LL}^0}  + { A_{LL}^{\cos \phi}} \,  \cos \phi   
\end{eqnarray}
    \end{subequations}
where,
\begin{eqnarray}
\label{eq:ALU2}
A_{LU}  & = & \frac{N^{\rightarrow} -N^{\leftarrow}}{N^{\rightarrow} + N^{\leftarrow}} = \sqrt{\epsilon(1-\epsilon)}  \, \Bigg[ \frac{F_{LU}^{\sin \phi} }{F_{UU,T} + \epsilon F_{UU,L}} \Bigg]\\
\label{AUL:eqn}
A_{UL} & = & \frac{N_{s_z=+}-N_{s_z=-}}{N_{s_z=+}+N_{s_z=-}} = \sqrt{\epsilon(\epsilon+1)} \Bigg[ \frac{F_{UL}^{\sin \phi}}{{F_{UU,T}   +\epsilon \, F_{UU,L} }} \Bigg] \sin \phi   +\epsilon \,  \Bigg[ \frac{ F_{UL}^{\sin 2 \phi}  }{F_{UU,T} +\epsilon \, F_{UU,L} } \Bigg] \sin 2 \phi \\
\label{eq:ALL2}
A_{LL} & = & \frac{N^\rightarrow_{s_z=+}-N^\rightarrow_{s_z=-}+ N^\leftarrow_{s_z=+}-N^\leftarrow_{s_z=-}}{N_{s_z=+}+N_{s_z=-}} = \sqrt{1 - \epsilon^2}  \, \Bigg[ \frac{F_{LL}}{F_{UU,T} + \epsilon F_{UU,L} } \Bigg]  +   \sqrt{\epsilon(1-\epsilon) } \, \Bigg[  
\frac{ F_{LL}^{\cos \phi}}{F_{UU,T} + \epsilon F_{UU,L}} \Bigg] \, \cos \phi \nonumber \\
\end{eqnarray}
in the equations, $N^{\rightarrow(\leftarrow)}_{s_z=\pm}$ describes a righthanded ($\rightarrow$), or lefthanded ($\leftarrow$) lepton scattering on a proton with longitudinal spin, $s_z= \pm 1/2$.  



In summary, based on the original work presented in Ref.\cite{Goldstein:2013gra}, we listed the expressions for the structure functions entering the observables for $\pi^o$ electroproduction from an unpolarized, longitudinally polarized and transversely polarized nucleon target, written in terms of the helicity amplitudes and CFFs. We work in a framework where the cross section is assumed to be dominated by all twist-two chiral-odd GPDs for transverse virtual photon polarization and by twist-two chiral even, axial vector GPDs for longitudnally polarized virtual photon. 
The observables highlighted in boldface have been measured at Jefferson Lab and they represent, to our knowledge, all available data for $\pi^0$ electroproduction, given as a function of $\xi,t, Q^2$.
The full data set is further described in Section \ref{subsec:DATA}.

\section{Likelihood Analysis}
\label{sec:MCMC}

Extracting the underlying CFFs from deeply virtual exclusive scattering observables constitutes a nonlinear inverse problem. In general, a set of (N) measured observables and (N) unknown parameters does not imply a unique solution when the observables depend nonlinearly on those parameters. The corresponding system of equations can admit multiple discrete solutions, continuous or nearly continuous degeneracies, or regions of parameter space that are only weakly constrained by the data. Moreover, experimental uncertainties transform these mathematical solutions into finite allowed regions whose geometry can be highly non-Gaussian.

This situation arises directly in deeply virtual $\pi^0$ electroproduction, where the measured differential cross sections and asymmetries are decomposed 
into response functions associated with different photon polarizations and interference terms as in Eqs.\eqref{eq:allosbervables}.
The measured quantities constrain quadratic combinations of the CFFs rather than the individual CFFs themselves. Different sets of CFFs can consequently produce the same, or experimentally indistinguishable, combinations of response functions.
The extraction should, therefore, not be viewed simply as the inversion of a system with an equal number of equations and unknowns, but it is about defining the regions of the multidimensional CFF parameter space that are consistent with the measurements. The likelihood-based Bayesian analysis adopted here provides a natural framework for this purpose, while MCMC sampling allows the full structure of the resulting probability distribution—including correlations, degeneracies, and multiple solution branches—to be explored.

\subsection{Experimental Data Used in this Analysis}
\label{subsec:DATA}
For this analysis, we have chosen experimental data on cross-section of Deep Exclusive Electroproduction of $\pi ^0$ at High $Q^2$ in the Quark Valence Regime from the Jefferson HALL A collaboration with the kinematic parameters $[E, x_{Bj}, Q^2, t, \epsilon]$ \cite{JeffersonLabHallA:2020dhq}. Another cross-section data that we have used is the Exclusive neutral pion electroproduction in the deeply virtual regime \cite{Fuchey:2010xsx} again from the Jefferson HALL A collaboration. Besides this there are some data available from HALL B on asymmetries ($\mathrm{A_{UL},A_{LL},A_{LU}}$) \cite{Kim:2015pkf, CLAS:2023wda}.  In addition, we incorporate a cross-section dataset from the CLAS (Hall B) collaboration into our likelihood analysis \cite{CLAS:2012cna, CLAS:2014jpc}. This specific dataset was selected for the in-depth study and for joint analysis with the asymmetry data due to its superior statistical power. We will elaborate more on the choice in the results sections as to how data on cross-sections from HALL B joined with data on asymmeteries from HALL B offers more robustness, higher quality, or yields stronger statistical power.

\subsection{Likelihood Analysis} 
\label{subsec:likeli}
Let us consider a random sample $\mathrm{X_1, X_2,\ldots,X_n}$ whose assumed probability distribution is dependent on some unknown parameter $\theta$. In such a scenario, our primary objective is to construct a point estimator u($\mathrm{X_1, X_2,...,X_n}$), such that  u($\mathrm{x_1, x_2,\ldots,x_n}$) is a good point estimate of $\theta$ where $\mathrm{x_1, x_2,\ldots,x_n}$ are the observed values (experimental data) of the random sample. For example, for a random sample $\mathrm{X_1, X_2,...,X_n}$ for which the $X_i$ are assumed to be normally distributed with mean $\mu$ and variance $\sigma^2$, then we should be able to find a good estimate of $\mu$ using the observations $\mathrm{x_1, x_2,...,x_n}$ that we obtained from our specific random sample. 
In other words, a good estimate of the unknown parameter $\theta$ would be the value of $\theta$ that maximizes the probability or the likelihood of getting the data we observed. This is in general the idea behind maximum likelihood estimation (amore in depth description is given {\it e.g.} in Ref.\cite{Hogg:2010yz}). In practice, assuming that the experimental measurements are statistically independent, the joint probability density is written in a factorized form as,
\begin{equation}
    \mathcal{L}(\theta) = P(X_1 = x_1, X_2 = x_2, \dots, X_n = x_n)
                = f(x_1; \theta) \cdot f(x_2; \theta) \cdots f(x_n; \theta)
                = \prod_{i=1}^{n} f(x_i; \theta)
                \label{eq:likeli}
\end{equation}
where $f(x_i;\theta)$ is the probability density function for the (i)-th observation, evaluated for a given set of model parameters ($\theta$).
The first equality is the definition of likelhood in terms of the joint probability density of the onbserved data.
The second equality follows from the assumption that the observations are statistically independent 
and the last equality is the shorthand notation of the product of the indexed terms, or the product of gaussians. 
If the individual measurements are assumed to follow Gaussian distributions, the likelihood consequently takes the form of a product of Gaussian probability densities.
Note that, while in the present analysis, correlations among different experimental measurements are neglected, this concerns only the experimental inputs to the likelihood and it should be distinguished from correlations among the model parameters. The latter will arise from the joint posterior probability distribution and constitute an important part of the present analysis (more details of this methodology as applied to deeply virtual scattering experimental data, can be found in our recent work in Ref.\cite{Adams:2024pxw}). 

For a fixed set of kinematic parameters $[E, x_{Bj}, Q^2, t]$, Eq.\eqref{eq:kinematics}, the cross-section for exclusive $\pi^0$ electroproduction can be expressed in terms of structure functions arising from the contributions of different helicity amplitudes asscociated with the initial and final particle helicities, Eq.\eqref{xs}. 

In this work, we use the cross section model from Refs.\cite{Goldstein:2013gra,Goloskokov:2011rd,Goloskokov:2012bf} center on chiral-odd GPDs,  
and data from \cite{JeffersonLabHallA:2020dhq,Kim:2015pkf,Bedlinskiy:2012be}. The standard deviation for each data point is taken as the total experimental uncertainty, with statistical and systematic uncertainties added in quadrature. The systematic uncertainties reported by the experiments correspond, on average, to approximately $4.5\%$) of the measured values and are smaller than the statistical uncertainties in most kinematic bins.
The difference between the model predictions and the observed data are assumed to follow independent Gaussian distributions with these standard deviations. Therefore, we use a univariate Gaussian likelihood for each data point to evaluate the probability of the observed data. This assumes statistically independent measurements and neglects correlations among experimental uncertainties (Eq.\eqref{eq:likeli}. For our model parameters, 
For the model parameters, we adopt uniform priors over a finite interval, initially chosen as (-100,100). The prior ranges are subsequently varied to assess the boundedness of the parameters and to ensure that the posterior distributions are not artificially constrained by the chosen prior boundaries. 
The full likelihood is given by taking the product over all data points. Bayes' theorem then gives the joint posterior distribution,
\begin{equation}
p(\boldsymbol{\theta}\mid D)
\propto
\mathcal{L}(D\mid\boldsymbol{\theta}),
\pi(\boldsymbol{\theta}),
\end{equation}
where (D) denotes the experimental data and ($\pi(\boldsymbol{\theta})$), the prior distribution.
The posterior distribution is subsequently sampled using the Markov Chain Monte Carlo (MCMC) algorithm. 
The generated samples provide a numerical representation of the joint posterior, allowing us to explore the multidimensional parameter space and determine parameter estimates, uncertainties, credible intervals, and correlations among the inferred parameters. 

\subsubsection{Markov Chain Monte Carlo (MCMC) Sampling}
MCMC sampling methods comprise a family of algorithms designed to sample from probability distributions efficiently. Given our current problem, we use 
the affine-invariant ensemble MCMC sampler, inspired by the Metropolis-Hastings algorithm, implemented in the \texttt{emcee} package \cite{Foreman_Mackey_2013}. 
{Using the the affine-invariant ensemble MCMC sampler provides a middle ground method between the standard Metropolis-Hastings which is easier to set up compared to the Hamiltonian MCMC which requires the gradients to be defined symbolically.}

The sampling is based on having a guided exploration for the problem, namely, the algorithm  suggests a direction -- proposal distribution -- but before the walker commits to that direction, it makes a decision whether it is worth pursuing -- acceptance-rejection step. 
The entire distribution can therefore be explored, even when direct sampling is not feasible. 
The generated samples provide a numerical representation of the posterior distribution and can be used to visualize its multidimensional structure, including marginal distributions and correlations among the model parameters. A more detailed discussion of the \texttt{emcee} ensemble sampling algorithm and its application to an inference problem closely related to the one considered here is provided in Ref.~\cite{Adams:2024pxw}.

\section{Results and Discussions}
\label{sec:results}
We performed a Bayesian analysis, MCMC sampling, of the currently available experimental observables for deeply virtual exclusive $\pi^0$ electroproduction with total sample run of $\mathrm{10^7}$. Here, we present the resulting extraction of the chiral-odd CFFs, including their posterior distributions, uncertainties, and correlations.  
These results provide the first fully quantitative constraints on the chiral-odd CFFs from the available data. This analysis allows us to assess both the sensitivity of the measured observables to the individual CFFs and the extent to which they can be disentangled within the present theoretical framework.

For each kinematic bin, an ensemble of $N_w = 144$ walkers for the helicity amplitudes (HA),  and and $N_w=60$  for the CFFs, is initialized uniformly within the bounded parameter domain defined by the prior ranges. 
The statistics reported above the 1D-histograms correspond to the posterior median and the central 68\% posterior credible interval, defined by the 16th, 50th, and 84th percentiles of the posterior samples for each parameter. The marker “Best Likelihood” in the figure denotes that the sampled parameter set maximizes the likelihood among the displayed MCMC samples. The “Likelihood Mean” marker denotes the likelihood-weighted average of the displayed MCMC samples. This is solely for visualization purpose and not used as the quoted parameter estimate. All subsequent results were produced from a combined set of final $\mathrm{20k}$ MCMC samples aggregated across all ensemble walkers.


The convergence was monitored by inspecting both the {\it autocorrelation time}, $\tau$, which measures the number of MCMC steps over which successive samples remain correlated, and the {\it rejection rate}  which measures the fraction of proposed moves that are rejected, as functions of the chain length of the walkers.
The estimated integrated autocorrelation times for the parameters ranges from approximately $\mathrm{(15-40)}$ steps at the maximum chain length $\mathrm{\sim 15k}$. The autocorrelation time provides a measure of the number of MCMC steps required to obtain effectively independent samples. 
An initial {\it burn-in} period is discarded to remove the transient part of the chains during which the walkers retain sensitivity to their initial conditions and have not yet reached the region representative of the stationary posterior distribution.
Our criterion for burn-in was to keep only the final 20,000 samples after checking convergence so that we have a sufficiently high number of samples for visualization and marginalization. 
Reliable sampling therefore requires chains that are substantially longer than $\tau$. In practice, a chain length of at least approximately (50$\tau$) provides a useful criterion for obtaining a reliable estimate of the autocorrelation time. The rejection rate provides a complementary measure of the sampling efficiency. 
The mean acceptance fraction of the ensemble was monitored and presented in the following subsections. 

An additional feature of the present analysis is a {\it posterior consistency check} in which samples drawn from the inferred parameter distributions are propagated back to the measured observables. This provides a direct test of whether the parameter regions identified by the Bayesian analysis can reproduce the experimental measurements within their quoted uncertainties. As shown below, the resulting distributions exhibit good agreement with the data, providing an important validation of the extracted, potentially multimodal, posterior distributions.

\begin{figure}[H]
    \centering
    \includegraphics[width=0.45\linewidth]{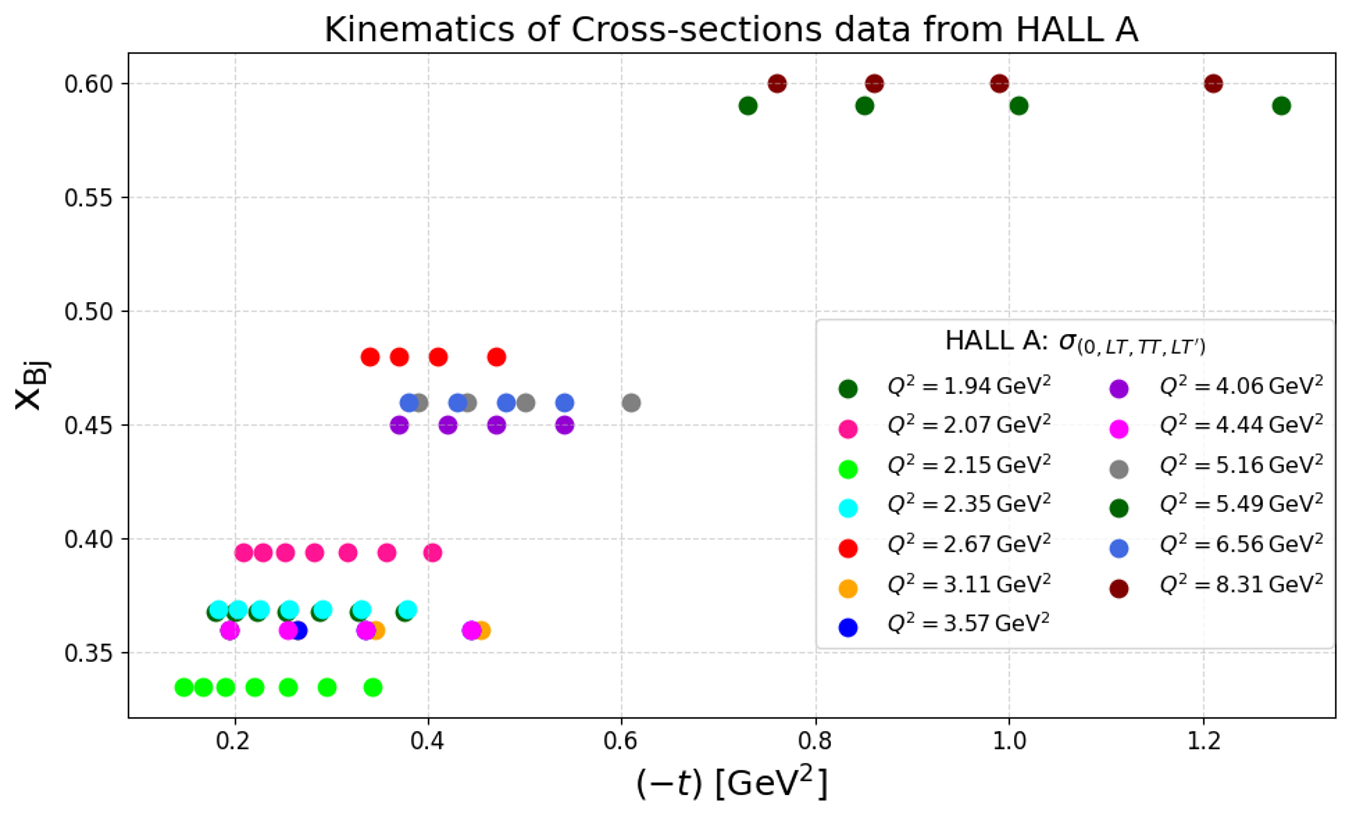}
    \hspace{0.35cm}
    \includegraphics[width=0.45\linewidth]{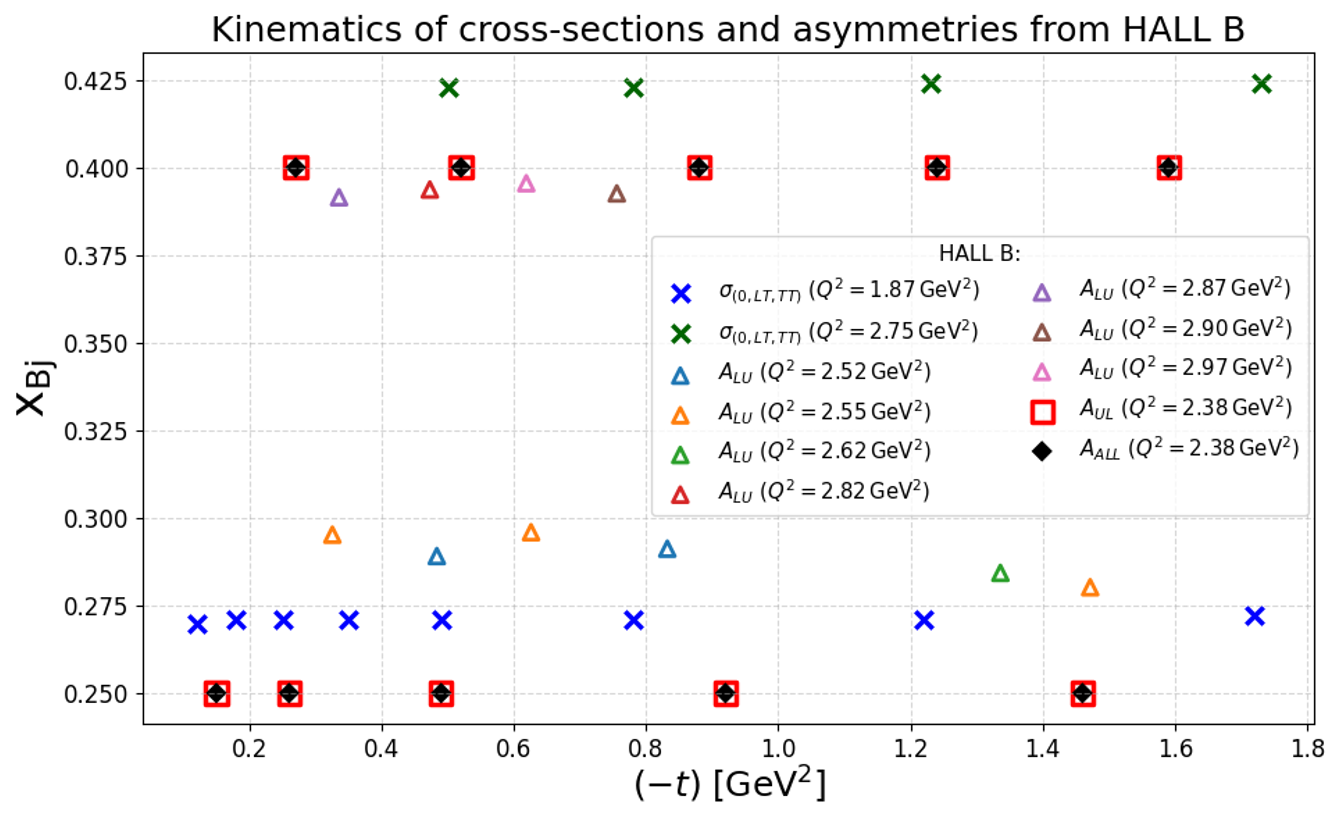}
    \caption{Experimental data from Jefferson Lab Hall A \cite{JeffersonLabHallA:2020dhq, Fuchey:2010xsx} and Hall B \cite{CLAS:2012cna, CLAS:2014jpc, Kim:2015pkf, CLAS:2023wda} which have been used in the Bayesian Likelihood analysis. Certain cross-section datasets are omitted from the plot because they reside in a disjoint region of the kinematic phase space, completely isolated from the coverage of the asymmetry data utilized in this joint analysis.}
    \label{fig:data}
\end{figure}


\begin{table}[htbp]
\centering
\begin{tabular}{|c|c|c|c|c|c|c|c|c|c|c|c|c|c|c|c|c|}
\hline
 $Q^2$ & $x_B$ & $\epsilon$ & $(-t)$ & $\sigma_0$ & ${stat}$ & ${sys}$ & $\sigma_{LT}$ & ${stat}$ & ${sys}$ & $\sigma_{TT}$ & ${stat}$ & ${sys}$ & $\sigma_{LT'}$ & ${stat}$ & ${sys}$ 
\\
\hline
 &  &  & &  &  &  &  &  & &  & &  &  & & 
\\
2.67 & 0.48 & 0.51 & 0.47 & 480.77 & $23.45$ & $37.63$ & -60.2 & $30.66$ & $2.11$ & -116.12 & $57.67$ & $4.06$ & 14.05 & $27.05$ & $0.49$ \\
 &  &  & &  &  &  &  &  & &  & &  &  & & 
\\
\hline
 &  &  & &  &  &  &  &  & &  & &  &  & & 
\\
1.87 & 0.271 & 0.598 & 0.49 & 259.5 & $7.30$ & $13.0$ & 25.1 & $7.20$ & $6.10$ & -167.0 & $19.0$ & $14.0$ & - & - & - \\
 &  &  & &  &  &  &  &  & &  & &  &  & & 
\\
\hline
\end{tabular}
\caption{Table of Kinematic bins and cross-section values along with their uncertainties from Jefferson Lab HALL A \text{\it(first row)} and HALL B \text{\it(second row)} experiment used for Bayesian likelihood analysis \cite{JeffersonLabHallA:2020dhq, Fuchey:2010xsx}. The notations \text{\it(stat, and sys)} are the statistical and systematic uncertainties, respectively. The kinematic variables $\mathrm{Q^2, (-t)}$ are given in $(\mathrm{GeV^2})$ and the cross-sections are in units of $\mathrm{nb/GeV^2}$.} 
\label{tab:sigma_HALLA_HALLB_data}
\end{table}


\begin{table}[htbp]
\centering
\begin{tabular}{|c|c|c|c|c|c|c|c|c|c|c|c|c|c|c|c|}
\hline
$Q^2$ & $x_B$ & $\epsilon$ & $-t$ & $A_{UL}^{\sin\phi}$ & ${stat}$ & ${sys}$ & $A_{UL}^{\sin 2\phi}$ & ${stat}$ & ${sys}$ &$A_{LL}^{const}$ & ${stat}$ & ${sys}$ & $A_{LL}^{\cos\phi}$ & ${stat}$ & ${sys}$ \\
\hline
 &  &  & &  &  &  &  &  & &  & &  &  & & 
\\
1.94 & 0.25 & 0.51 & 0.49 & 0.292 & 0.036 & 0.021 & -0.073 & 0.036 & 0.031 &   0.564 & 0.036 & 0.031 & 0.2 & 0.064 & 0.078\\
 &  &  & &  &  &  &  &  & &  & &  &  & & 
\\
\hline
\end{tabular}
\caption{Table of Kinematic bins and Spin Asymmetry $\mathrm{A_{UL} \text{and} A_{LL}}$ values along with their uncertainties from Jefferson Lab HALL B experiment used for Bayesian likelihood analysis \cite{Kim:2015pkf, CLAS:2023wda}.The notations \text{\it(stat, and sys)} are the statistical and systematic uncertainties, respectively. The kinematic variables $\mathrm{Q^2, (-t)}$ are given in $(\mathrm{GeV^2})$.}
\label{tab:asym_data}
\end{table}

\subsection{Joint Analysis of HALL A and HALL B experimental data}
In order to extract the helicity amplitudes or the chiral-odd Compton form factors from the available experimental data, Figure \ref{fig:data}, Table \ref{tab:sigma_HALLA_HALLB_data} and Tables \ref{tab:asym_data}, we use a joint likelihood analysis method as described in above section. Given the set of experimental data in the references \cite{JeffersonLabHallA:2020dhq, Kim:2015pkf, CLAS:2023wda}, selecting optimal kinematic bins for extracting helicity amplitudes (HAs) and Compton form factors (CFFs) from cross-section and asymmetry data is challenging, as the respective kinematic coverages do not share a common overlap. Looking closely at the experimental data available from the HALL A \cite{JeffersonLabHallA:2020dhq} and HALL B \cite{Kim:2015pkf, CLAS:2023wda}, we find that for a given kinematic bin $[E, x_{Bj}, Q^2, t, \epsilon]$, there is only one data point for each of the structure functions and asymmetries. Therefore, to optimize the utility of the available datasets, we minimize the kinematic discrepancy by calculating the Euclidean distance between each of the rows in the available data for the cross-sections and the asymmetries ($\mathrm{A_{UL}, A_{LL}}$). The specific pair for which the distance metric is minimum, are the two most closest matching kinematic bins across the datasets. The average of these two bins is used for the analysis. Of course the analysis can be performed for each set of kinematic bin but we defer this for future work. These closest matching kinematic bins obtained are mentioned in first row of Table \ref{tab:sigma_HALLA_HALLB_data} and Tables \ref{tab:asym_data}. 

\begin{figure}
    \centering
    \includegraphics[width=0.85\linewidth]{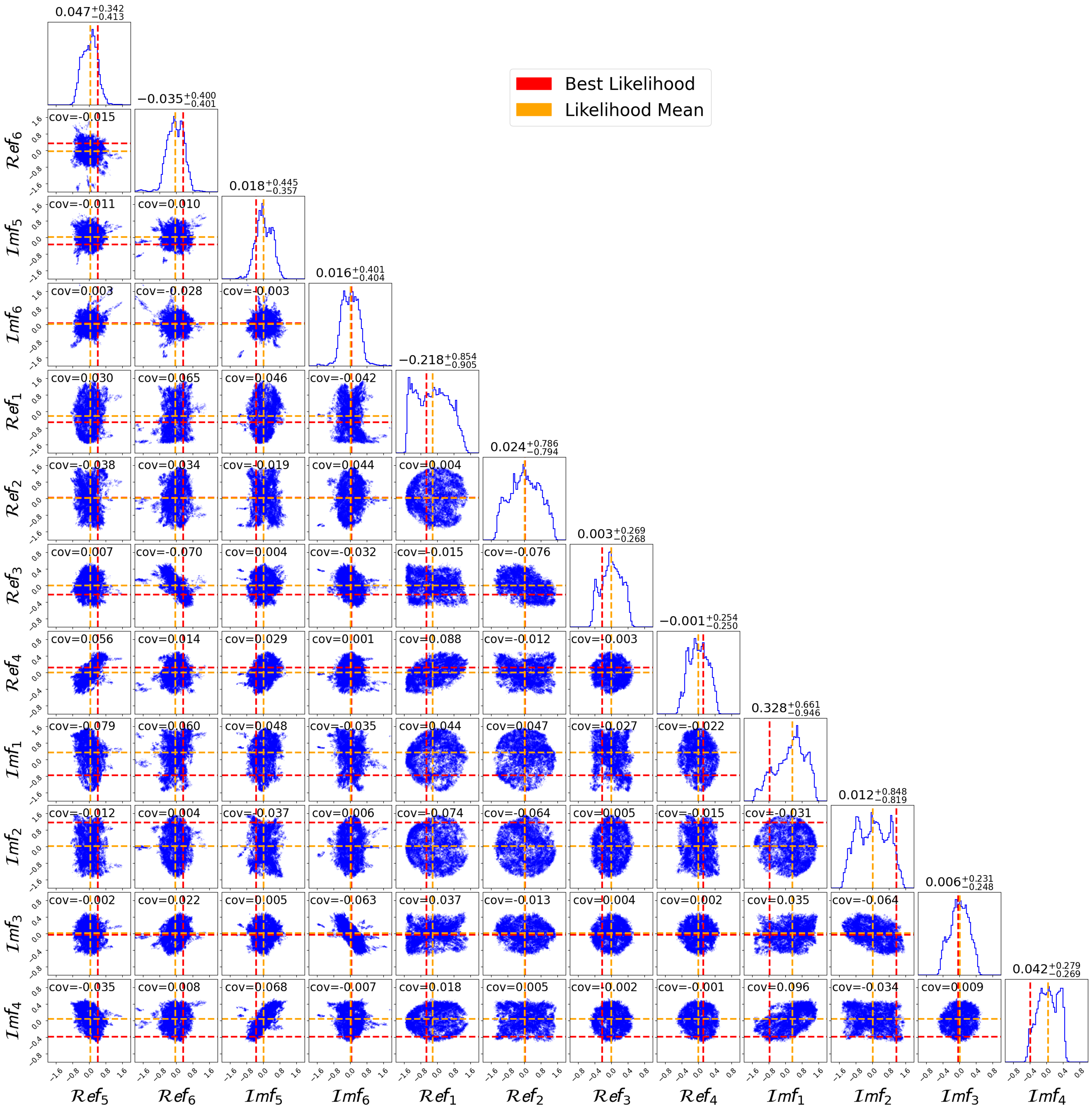}
    \caption{Extraction results for the chiral-odd helicity amplitudes from the joint likelihood analysis of Hall A and Hall B data. The chiral-even amplitudes are bounded within a uniform prior domain of $\mathrm{[-10,10]}$. The amplitudes described in terms of chiral even GPDs, $f_5$ and$f_6$ are constrained within a uniform prior domain of $\mathrm{[-10,10]}$.}
    \label{fig:HALLA_HALLB_joint}
\end{figure}

\begin{figure}
    \centering
    \includegraphics[width=0.85\linewidth]{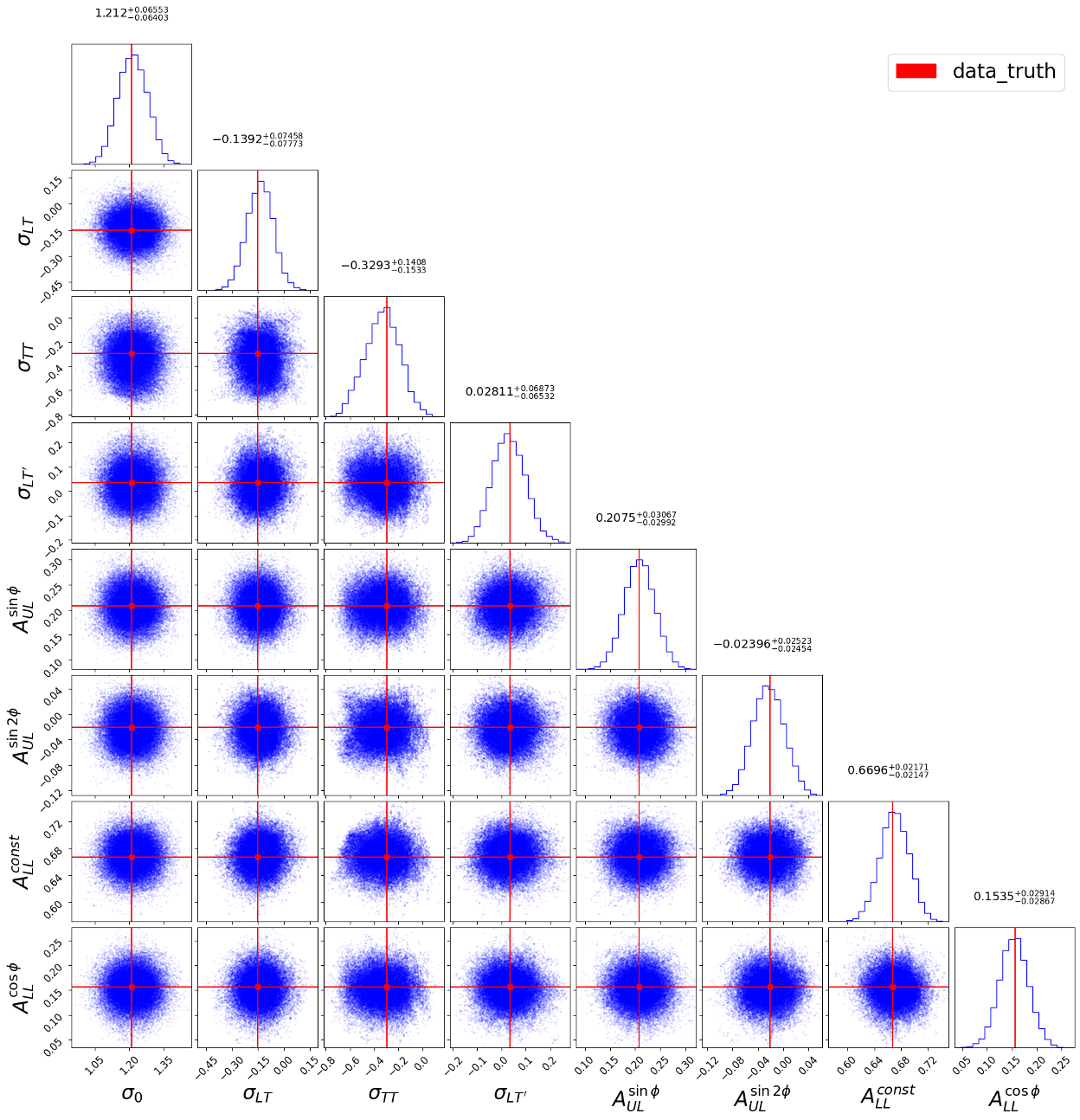}
    \caption{Consistency check on the experimental observables with the extracted chiral-odd helicity amplitudes using joint likelihood analysis of Hall A and Hall B data presented in figure \ref{fig:HALLA_HALLB_joint}. The amplitudes for chiral-even  GPDs, $f_5$ and $f_6$, are constrained within a uniform prior domain of $\mathrm{[-10,10]}$.}
    \label{fig:cc_HALLA_HALLB_joint}
\end{figure}

In Figure \ref{fig:HALLA_HALLB_joint}, we present a joint likelihood analysis on the HALL A cross-sections and HALL B asymmtery data. The MCMC samples obtained are presented in the form of corner plot. In the joint analysis, chiral-odd amplitudes are allowed to vary in the prior domain of $\mathrm{[-100,100]}$. The remaining chiral-even amplitudes are fixed in comparison to the other chiral-odd amplitudes varing them between $\mathrm{[-3,3]}$ only, motivated by physical constraints \cite{Goldstein:2013gra}. The MCMC samples presented in the form of corner plot show that the  parameters of the helicity amplitudes have high covariance, meaning highly correlated or anti-correlated. This implies that the experimental data fails to constrain the helicity amplitudes (parameters) independently. Besides this, we also find that the 1-D marginal distributions of the parameters are heavily multi-modal. This results in producing heavily misleading large values of the parameters even when prior domain is good enough. 


Figure \ref{fig:cc_HALLA_HALLB_joint} shows the consistency check or predictive power of the maximum likelihood analysis where the final
20000 MCMC samples are compared with the experimental measurements. The MCMC samples show a good agreement with the experimental data indicating that the extracted parameters distributions reproduce the experimental data within the quoted experimental uncertainties. This combination of multi-modal posterior distributions and strong predictive power indicates the under-constrained nature of the CFFs from experiments with inherent mathematical symmetries. The extracted modes in the corner plot of Figure \ref{fig:HALLA_HALLB_joint} are degenerate solutions that reconstruct the experimental data with equal precision. This indicates that non-linear structure of the underlying physics problem. From an experimental standpoint, extracting physics observables by combining datasets from distinct detector frameworks presents severe challenges, primarily due to non-overlapping kinematic coverages and differing systematic acceptances. This motivated us to do joint likelihood analysis on the experimental datasets produced from one detector framework, which might possibly help us solve some of the problems faced here. 

\subsection{Joint Analysis from HALL B experimental data only}

To achieve a more robust statistical scenario, we incorporate additional cross-section data from the CLAS detector in Hall B \cite{CLAS:2012cna, CLAS:2014jpc}. By restricting our analysis to a single experimental framework rather than a heterogeneous combination of detectors, we eliminate the cross-calibration discrepancies and systematic uncertainties inherited in mixed-framework datasets as we showed above. The choice of the kinematic bins for analysis was based on having matching t-values and at the same time be very close in $[E, x_{Bj}, Q^2, t, \epsilon]$. The closest kinematic bins found are mentioned in second row of Table \ref{tab:sigma_HALLA_HALLB_data} and and Table \ref{tab:asym_data}. In similar way other sets of kinematic bins were chosen for analysis in $(-t)$. which we will present in the coming subsection.

\begin{figure}
    \centering
    \includegraphics[width=0.85\linewidth]{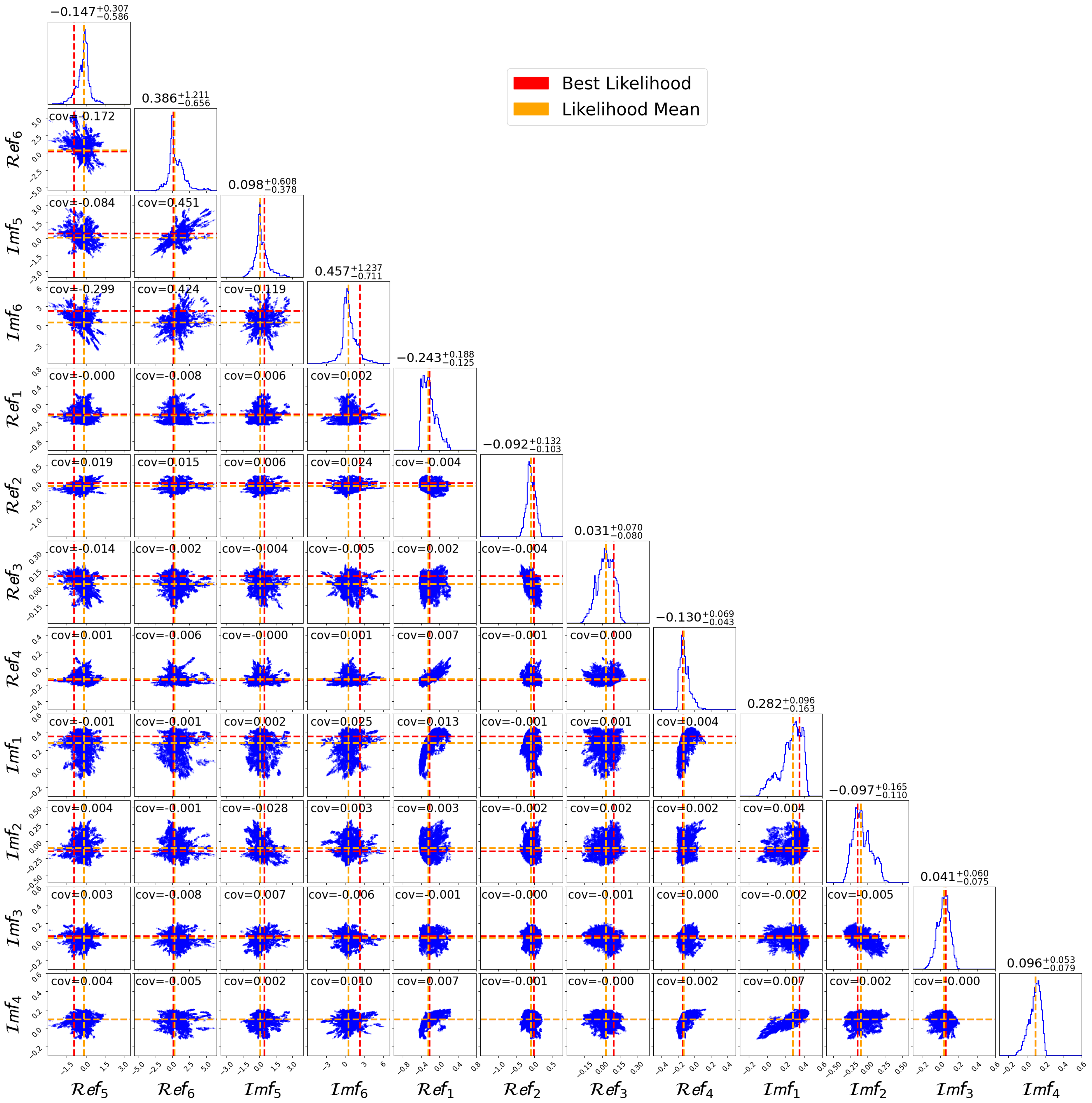}
    \caption{Extraction results for the chiral-odd helicity amplitudes from the joint likelihood analysis of only Hall B data with no inclusion of Beam Spin Asymmetry $\mathrm{A_{LU}}$. The accompanying chiral-even amplitudes are bounded within a uniform prior domain of $\mathrm{[-10,10]}$.}
    \label{fig:only_HALLB_joint}
\end{figure}

\begin{figure}
    \centering
    \includegraphics[width=0.85\linewidth]{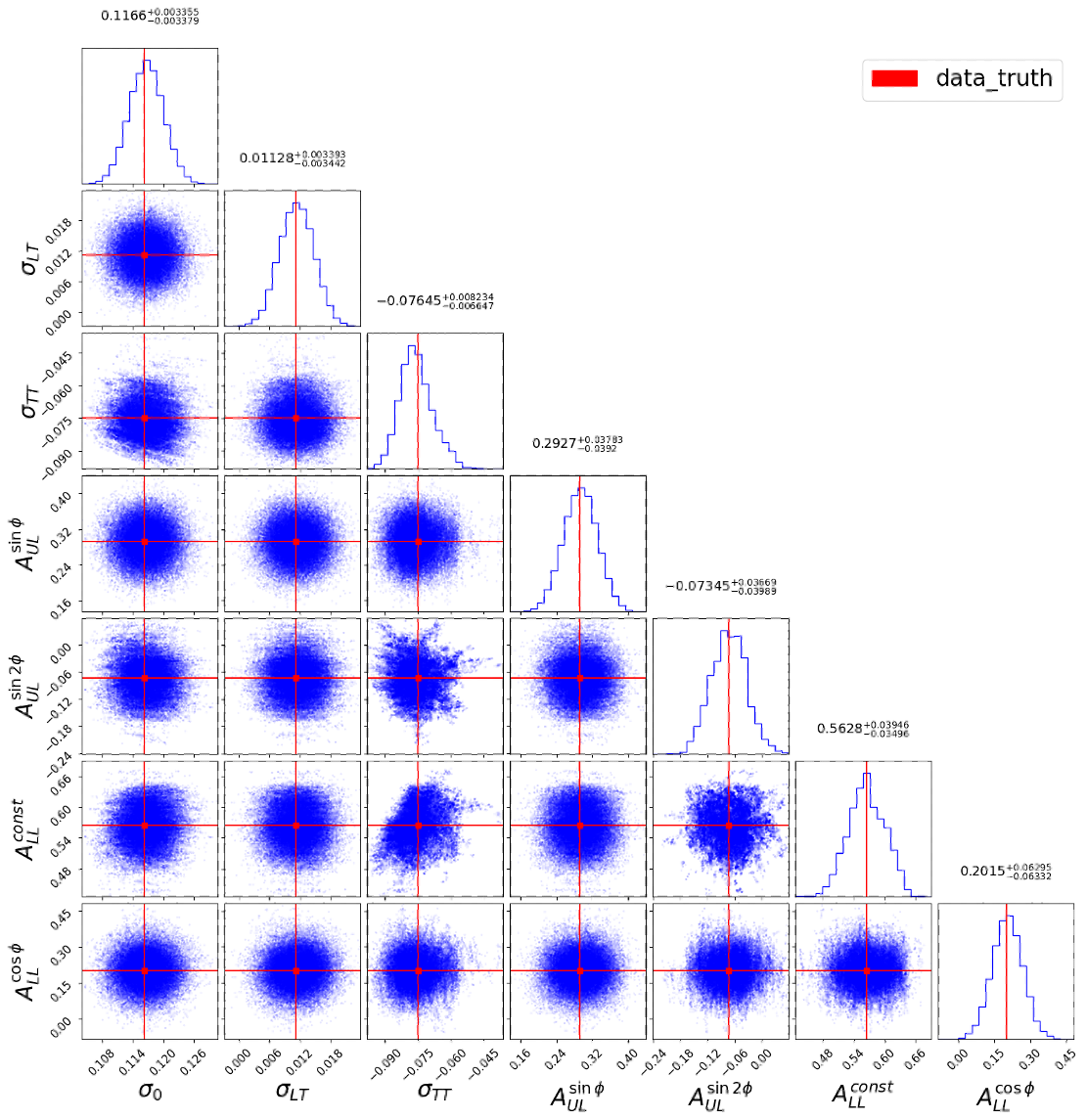}
    \caption{Consistency check on the experimental observables with the extracted chiral-odd helicity amplitudes using joint likelihood analysis of alone Hall B data without Beam Spin Asymmetry $\mathrm{A_{LU}}$ presented in Figure \ref{fig:only_HALLB_joint}. The accompanying chiral-even amplitudes are constrained within a uniform prior domain of $\mathrm{[-10,10]}$.}
    \label{fig:cc_only_HALLB_joint}
\end{figure}

Using the experimental data solely from HALL B collaboration, we did a joint likelihood analysis using again the cross-section and asymmetry values from the experimental data. It is to be noted that here, in case of HALL B data analysis, we do not have experimental data for the structure function $\mathrm{F^{\sin \phi}_{LU}}$ where the beam is longitudinally polarized and target is unpolarized. In Figure \ref{fig:only_HALLB_joint}, we show only HALL B data joint likelihood analysis. The procedure is similar to what we do for the mixed-framework dataset analysis, the difference only arising in the dataset and its source. Unlike the former scenario, here the MCMC samples shown in the form of corner plot appear to have smaller or no covariance. This means that the helicity amplitudes are not correlated or anti-correlated. The experimental data successfully constrains the helicity amplitudes independently. Besides this the 1-D marginal distributions are not multi-modal anymore generating values with suiTable prior domain.


Figure \ref{fig:only_HALLB_joint} shows the predictive power of the joint likelihood analysis from HALL B experimental data using final $20000$ MCMC samples. These MCMC samples reproduce experimental data within experimental uncertainties. This strong predictive power coupled with an albeit unimodal nature of posterior distributions, demonstrates that the helicity amplitudes can be robustly constrained under the premise of a correct model framework and consistent experimental data. 

\begin{figure}
    \centering
    \includegraphics[width=0.85\linewidth]{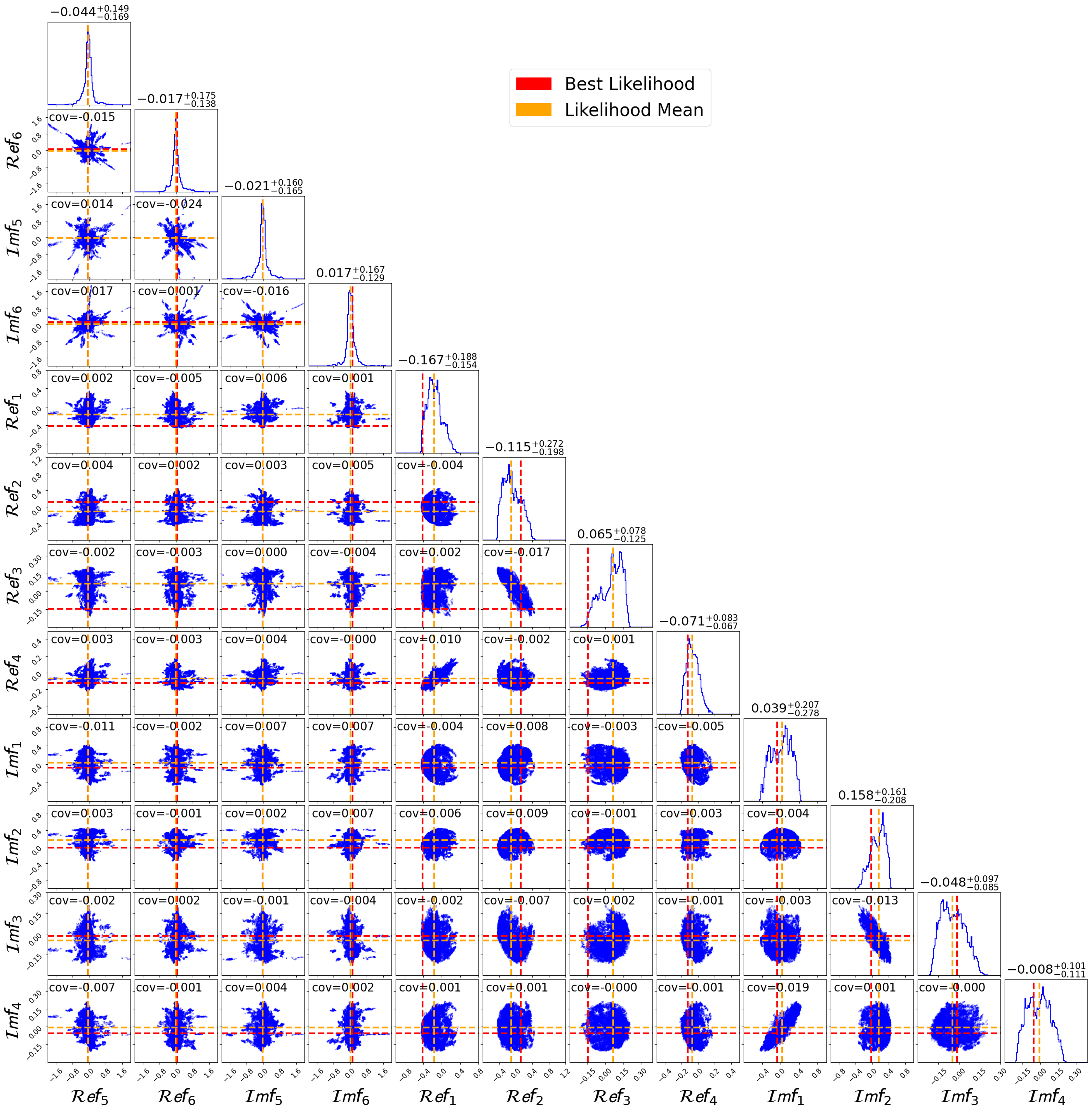}
    \caption{Extraction results for the chiral-odd helicity amplitudes from the joint likelihood analysis of Hall B data including, $\mathrm{A_{UL}, A_{LL}}$, with $\mathrm{A_{LU}}$. The accompanying chiral-even amplitudes are bounded within a uniform prior domain of $\mathrm{[-10,10]}$.}
    \label{fig:only_HALLB_joint_withALU}
\end{figure}

\begin{figure}
    \centering
    \includegraphics[width=0.85\linewidth]{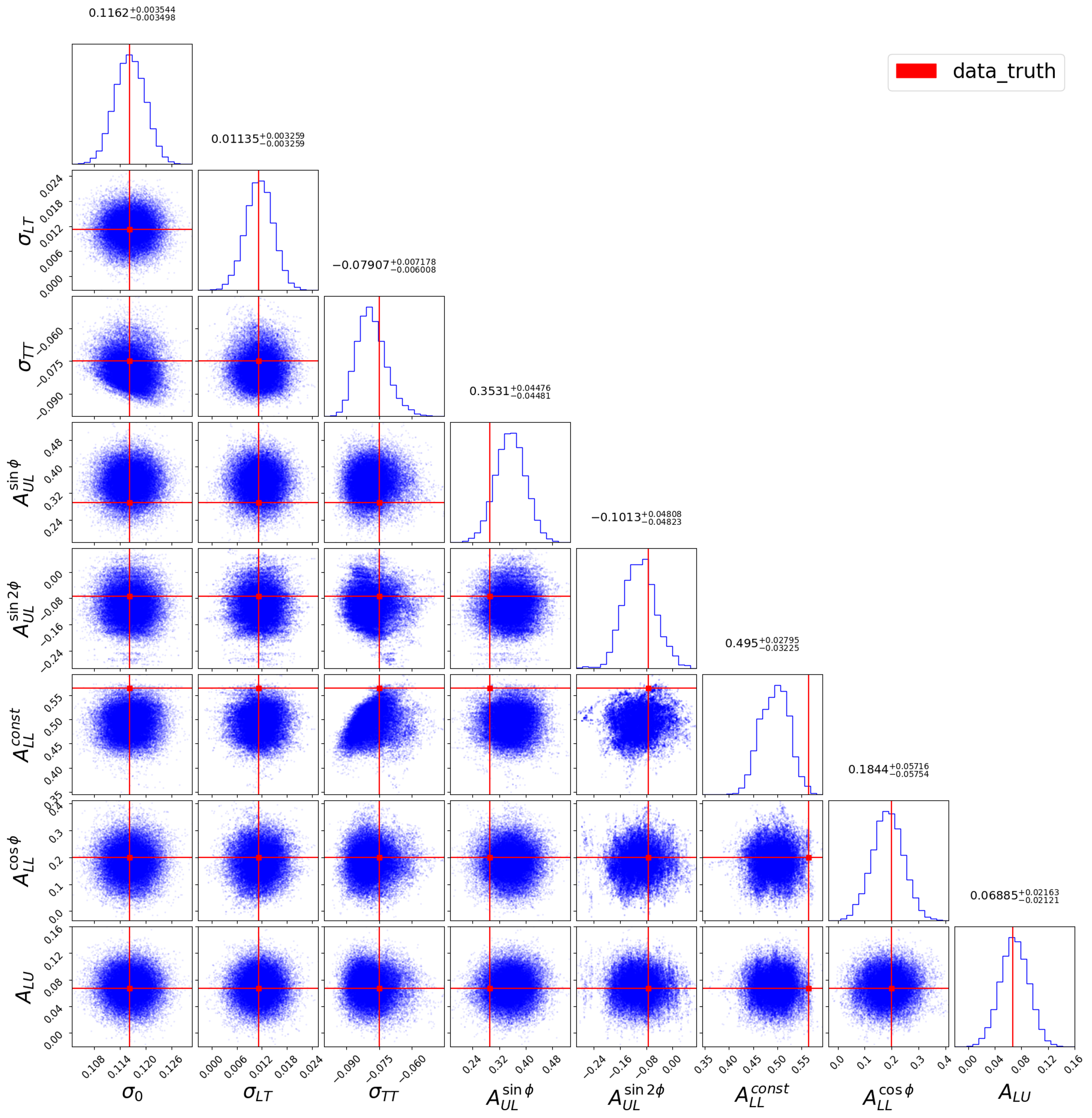}
    \caption{Consistency check on the experimental observables with the extracted chiral-odd helicity amplitudes using joint likelihood analysis of alone Hall B data including , $\mathrm{A_{UL}, A_{LL}}$, with Beam Spin Asymmetry $\mathrm{A_{LU}}$ presented in Figure \ref{fig:only_HALLB_joint_withALU}. The accompanying chiral-even amplitudes are constrained within a uniform prior domain of $\mathrm{[-10,10]}$.}
    \label{fig:cc_only_HALLB_joint_withALU}
\end{figure}

An additional aspect of this analysis involves verifying the statistical compatibility of the inclusion of Beam Spin Asymmetry (BSA) ($\mathrm{A_{LU}}$) in the likelihood analysis \cite{CLAS:2023wda}. We assess this by monitoring the stability of the posterior distributions and utilizing a posterior predictive check (PPC) or the consistency check when including the experimental data on ($\mathrm{A_{LU}}$). This is shown in Figures \ref{fig:only_HALLB_joint_withALU}. The inclusion of BSA in the likelihood appears to deteriorate the posterior distributions. even though the covariance is small but it increases the correlations compared to the scenario where we have no BSA in the analysis \ref{fig:only_HALLB_joint}. Besides this, the 1-D marginal distributions bring back the multi-modal nature of the chiral-odd helicity amplitudes producing misleading values of the parameters (HAs). 


The posterior predictive check on the MCMC samples deteriorates significantly upon the inclusion of the beam-spin asymmetry $(\mathrm{A_{LU}})$ data as shown in Figure \ref{fig:cc_only_HALLB_joint_withALU}. The final $20000$ MCMC samples exhibit poor fidelity and fail to satisfactorily reproduce the experimental data within their uncertainties. This stark tension arising when including experimental data on $(\mathrm{A_{LU}})$ indicates that the joint analysis with $(\mathrm{A_{LU}})$ introduces mutually incompatible constraints, rendering the extracted helicity amplitudes unphysical. However, we speculate that including BSA data might have made the joint analysis better if we had data on cross-section $\sigma_{LT'}$ from HALL B.

\subsubsection{Extraction of chiral-odd Compton form factors from Joint Likelihood Analysis using only HALL B data.}

\begin{figure}
    \centering
    \includegraphics[width=0.85\linewidth]{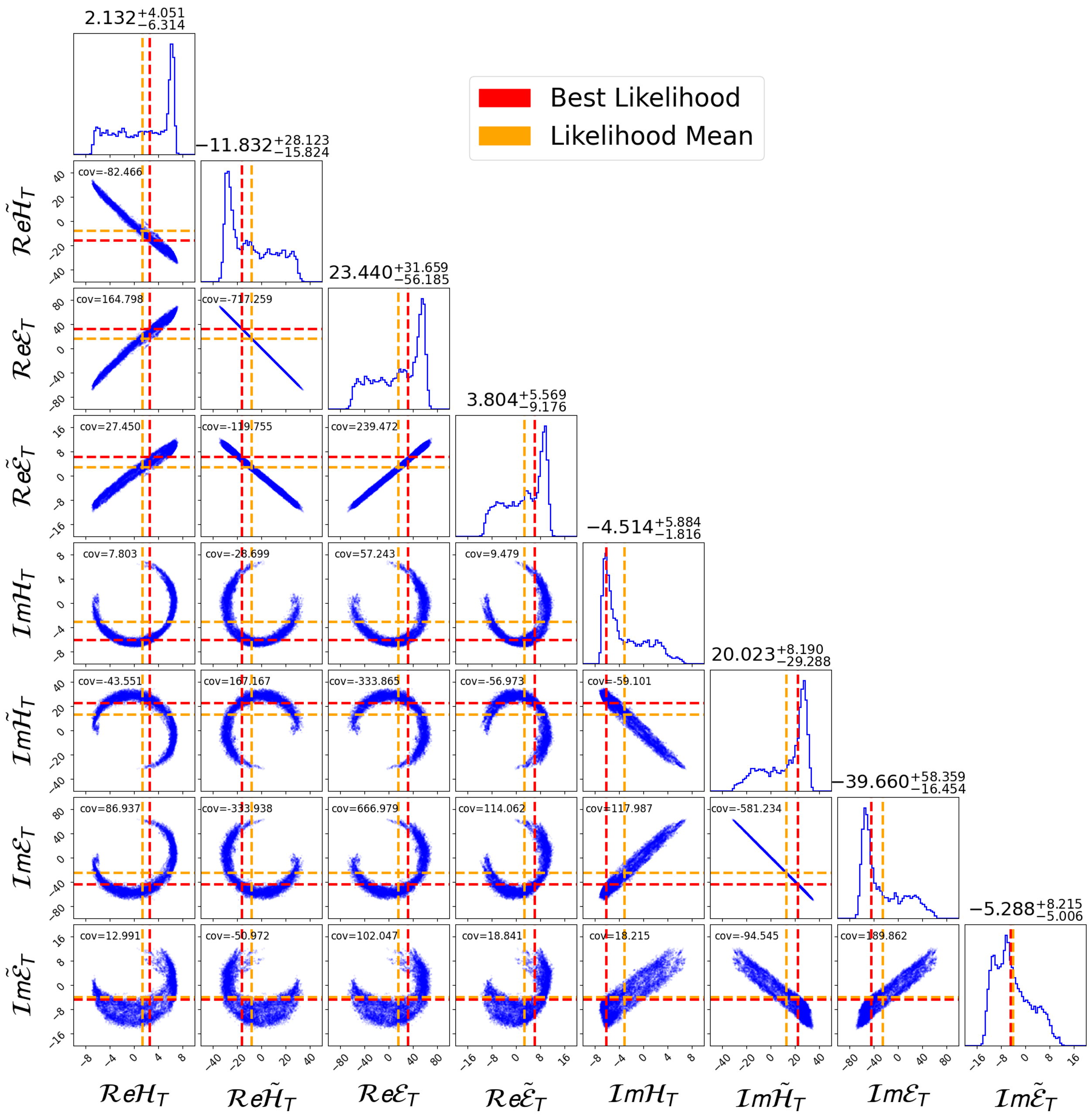}
    \caption{Extraction results for the chiral-odd Compton form factors from the joint likelihood analysis of only Hall B data without experimental data from HALL B on Beam Spin Asymmetry $\mathrm{A_{LU}}$. The accompanying chiral-even amplitudes are bounded within a uniform prior domain of $\mathrm{[-10,10]}$.}
    \label{fig:only_HALLB_joint_CFFs}
\end{figure}

\begin{figure}
    \centering
    \includegraphics[width=0.85\linewidth]{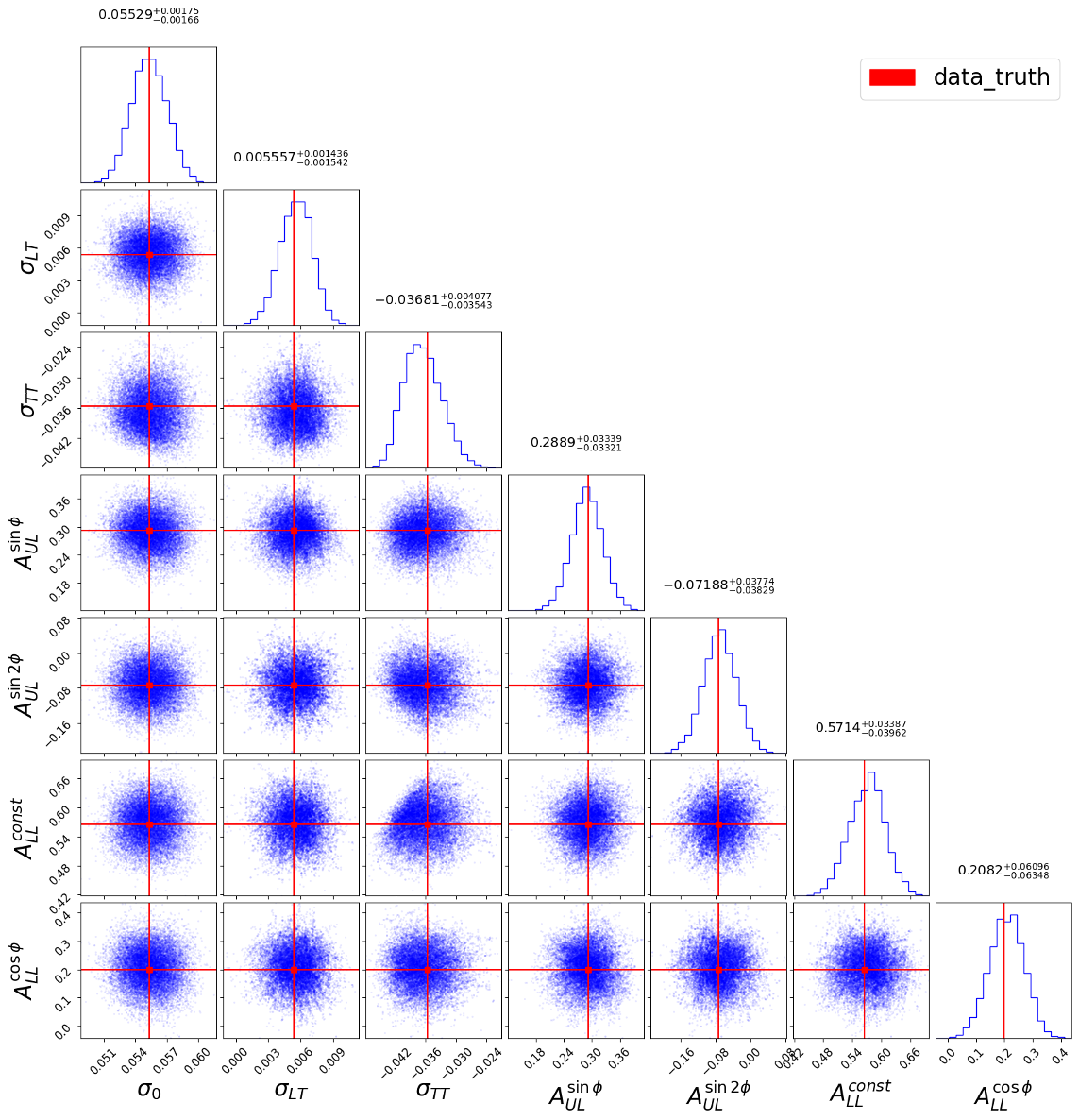}
    \caption{Consistency check on the experimental observables with the extracted chiral-odd Compton form factors using joint likelihood analysis of alone Hall B data without Beam Spin Asymmetry $\mathrm{A_{LU}}$ presented in Figure \ref{fig:only_HALLB_joint_CFFs}. The accompanying chiral-even amplitudes are constrained within a uniform prior domain of $\mathrm{[-10,10]}$.}
    \label{fig:cc_only_HALLB_joint_CFFs}
\end{figure}

Using Jefferson Lab HALL B experimental data on cross-sections from the CLAS detector \cite{CLAS:2012cna, CLAS:2014jpc} and asymmetries ($\mathrm{A_{UL}, A_{LL}}$) \cite{Kim:2015pkf, CLAS:2023wda}, we do a joint likelihood analysis on chiral-odd Compton-Form Factors independently. The kinematic bin configurations chosen for this joint analysis on CFFs are identical to those utilized in the joint likelihood analysis of the helicity amplitudes. Initially, we adopted a uninformed prior bound of [-100, 100] across the parameters. A careful review of the MCMC chains showed that the high-probability regions occupied a significantly smaller volume of the parameter space. Therefore, we updated the model with parameter (CFFs) specific narrower prior domains while keeping the converged posteriors unaffected, thereby ensuring the robustness of parameter estimation. The final MCMC samples of the extracted chiral-odd Compton form factors along with their consistency check are presented in Figure \ref{fig:only_HALLB_joint_CFFs}. The predictive power of the analysis can be seen in the right panel of Figure \ref{fig:cc_only_HALLB_joint_CFFs} which again shows good agreement with the experimental data. A detailed explanation on Consistency check and why it is needed is mentioned in the next subsection.


\subsection{Consistency check and QCD Sanity check}
\textit{Significance of Consistency check or Posterior Predictive Check (PPC):} A Posterior Predictive Check (PPC) is the definitive Bayesian tool for evaluating whether a statistical or physical model is consistent with real experimental information or not. A MCMC run only ensures that it found the best possible parameters for the model one proposed. But it does not tell us if the model was wrong to begin with. A PPC ensures this by defocusing us from the abstract parameter space and asking the model to generate synthetic data (data from MCMC samples) which now we can compare directly with the real experimental data. If a model truly fits the data, then the recovered data generated by plugging in the MCMC samples into the model should agree with real experimental data. In this way, a PPC helps us diagnose the model and evaluate its predictive power. It is a two-step process: (1) Use my most likeli MCMC samples of parameters and pass them through our model to generate an entirely new dataset. (2) Plot these generated dataset as a matrix of distributions in the form of corner plot where one can visualize the multi-dimensional correlations of observables against single experimental data-point. \\

\begin{figure}[htbp]
    \centering
    \begin{subfigure}[b]{0.45\linewidth}
        \centering
        \includegraphics[width=\linewidth]{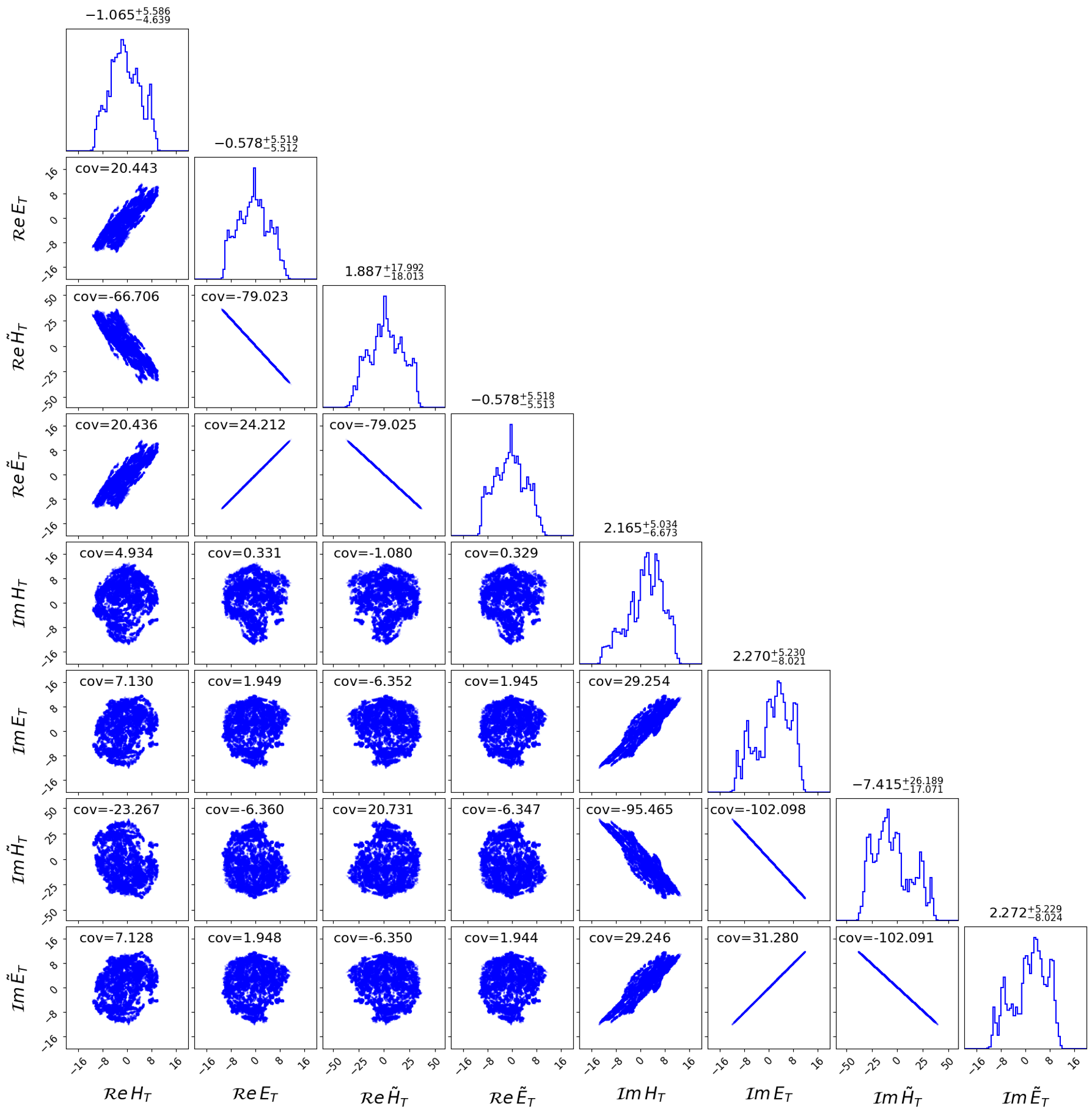}
        \caption{\it Forward extraction (HAs to CFFs)}
        \label{fig:sanity_check_has_to_cffs}
    \end{subfigure}
    \hfill 
    \begin{subfigure}[b]{0.45\linewidth}
        \centering
        \includegraphics[width=\linewidth]{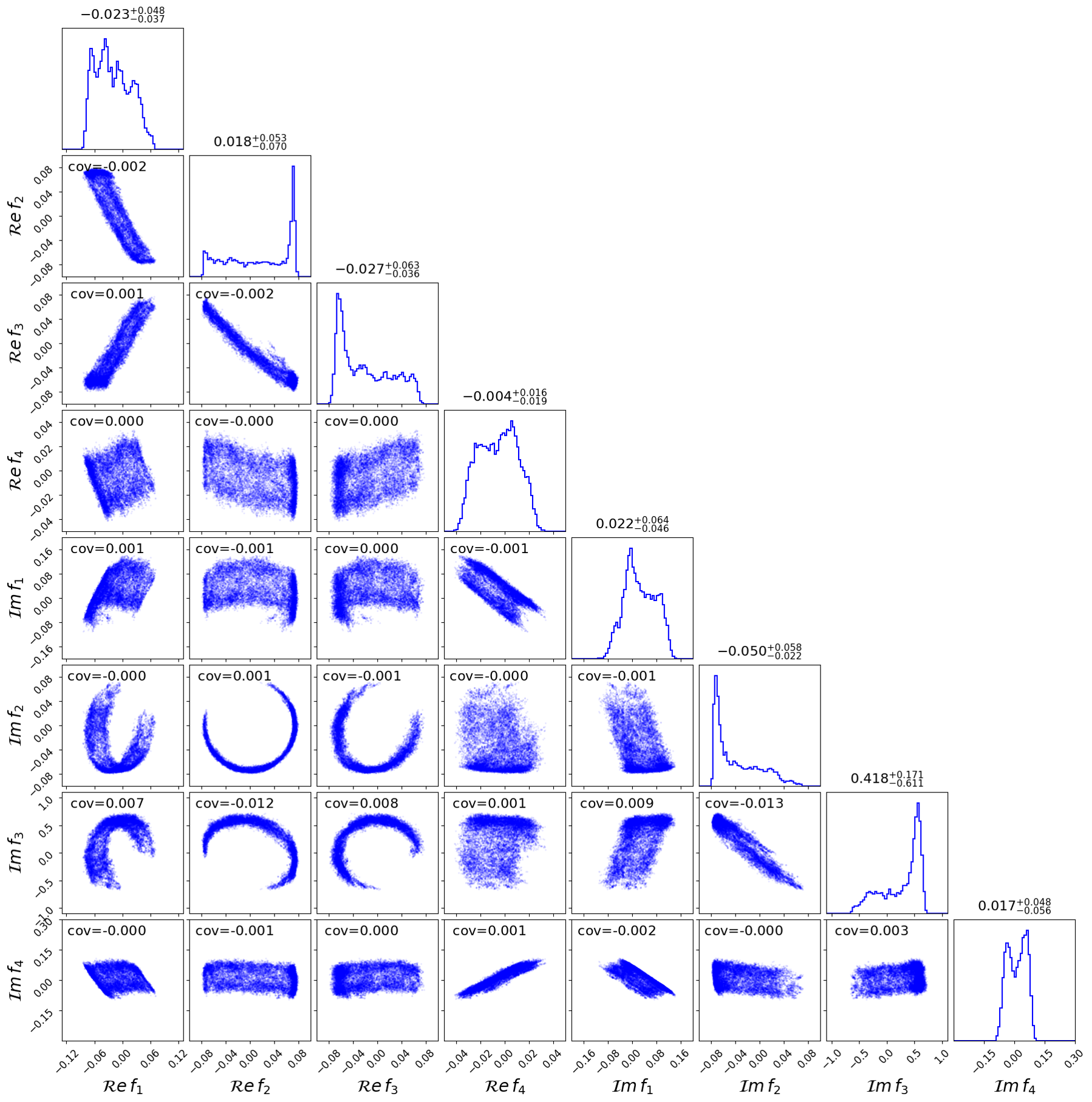}
        \caption{\it Reverse extraction (CFFs to HAs)}
        \label{fig:sanity_check_cffs_to_has}
    \end{subfigure}
    \caption{{\it (a)}: Sanity Check on the extracted chiral-odd helicity amplitudes from the joint likelihood analysis of only Hall B data. The MCMC is run on the helicity amplitudes independently and the chiral-odd CFFs are calculated using the mathematical expressions obtaining results similar to what we have in Figure \ref{fig:only_HALLB_joint_CFFs}. {\it (b)}: Sanity Check on the chiral-odd Compton form factors extracted from the joint likelihood analysis of only Hall B data. The MCMC is run on the chiral-odd Compton form factors independently and the chiral-odd helicity amplitudes are calculated using the inverted mathematical expressions producing results similar to that in Figure \ref{fig:only_HALLB_joint}. {\it Both}: The accompanying chiral-even amplitudes are constrained within a uniform prior domain of $\mathrm{[-10,10]}$ and the analysis does not include data on Beam Spin Asymmetry $\mathrm{A_{LU}}$.}
    \label{fig:sanity_checks}
\end{figure}

\textit{Significance of Sanity Check:} 
In context of Bayesian statistics, a deeper check that follows is the sanity check or Basis Independence test. This tells us that our extraction of (HAs or CFFs) parameters is independent of the basis we choose to sample in. As a sanity check for model invariance for both helicity amplitudes and Compton form factors, we perform physics extraction using two separate basis configurations: (1) We first run our emcee on our observables with helicity amplitudes as parameters. Use the obtained MCMC samples to calculate the Compton form factors. (2) We run a separate emcee on the CFFs and extract them as parameters. Use these parameters to obtain the helicity amplitudes through the inverted equations. By doing this we determine whether information is lost depending on which parameter space is chosen as the fundamental starting point for the analysis. Performing this forward and reverse extraction, we ensure that the choice of priors does not bias or drive the results we obtain. Both processes have their own consistency checks tested. The kind of solutions obtained will guide us as to what direction is more suiTable for the problem. This can be seen in Figure \ref{fig:sanity_checks}. Figure \ref{fig:sanity_check_has_to_cffs} shows the forward extraction of CFFs when the emcee is ran on helicity amplitudes and Figure \ref{fig:sanity_check_cffs_to_has} shows the reverse extraction of helicity amplitudes when emcee is performed on CFFs. Both scenarios agree well with each other within uncertainties. This tells us that our physics extraction is robust and unbiased by our choice of sampling variables. 

In the forward extraction (HAs to CFFs), the 1-D histograms have a very sharp, completely distinct double peaks or fluctuating multi-peaks but the 2-D plots are mostly uniform, densely filled blocks. When MCMC ran on helicity amplitudes, the sampler easily filled out the allowed space (Figure \ref{fig:only_HALLB_joint}), but when projected into CFF space, the non-linearities caused the probabilities to stack up into sharp, multi-peaked ridges. However, in reverse Extraction (CFFs to HAs), the 1-D histograms are mostly unimodal but the 2-D plots are either sharp diagonal lines or open ring arcs representing strong correlations. Although the covariance is almost negligible but it is difficult to find out the unique value of the amplitudes. Another drawback for the reverse extraction is that running an MCMC on CFFs comes with high computational costs. The sampler run for reverse extraction is almost (18-20) times greater than the sampler run for forward extraction. On the other hand, in the forward extraction, the sampler navigates and explores the allowed probability volume efficiently at the same time reveals the model degeneracies in a more transparent way which is unlikely to conclude from the reverse extraction. Thus, the forward extraction is a more efficient way to proceed to obtain a solution for the first inverse problem in the field of GPDs. This supports the paradigm that helicity amplitudes provide the primary basis for cross-sections, offering a more fundamental description than Compton form factors (CFFs). A more rigorous validation can be performed to confirm this but above evidences are enough to proceed. Thus, we follow the forward extraction for our analysis for the rest part of the paper.

\begin{figure}
    \centering
    \includegraphics[width=0.5\linewidth]{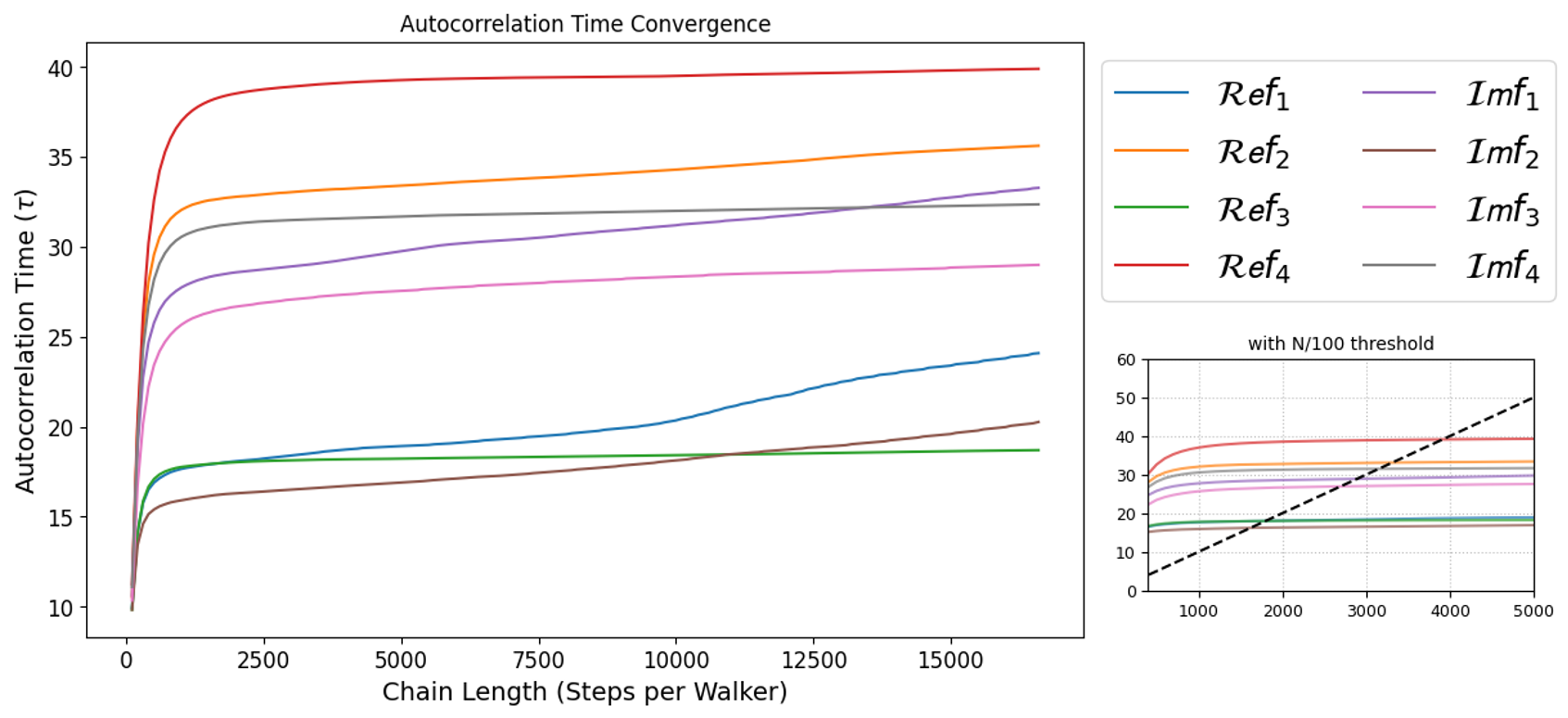}
    \caption{Convergence of the integrated auto-correlation time $\mathrm{\tau}$ as a function of MCMC chain length (steps per walker) for chiral-odd helicity amplitudes, with chiral-even amplitudes bounded in the prior domain $[-10, 10]$. The main plot illustrates the long-term stabilization of $\mathrm{\tau}$ across walker steps for the various components of the helicity amplitudes during a single MCMC run. The bottom-right inset highlights the early-stage sampling behavior alongside the standard $\mathrm{(N/100)}$ convergence threshold (dashed line). Correlations correspond to the helicity amplitude MCMC-chain run used to produce samples generated in Figure \protect\ref{fig:only_HALLB_joint}.}
    \label{fig:tau_all4}
\end{figure}

\subsection{Mean Acceptance Fraction and Auto-Correlation time}

In emcee, the integrated autocorrelation time $(\tau)$ is another premier diagnostic tool for quantifying the statistical efficiency and convergence of an ensemble Markov Chain Monte Carlo analysis. The consecutive steps in a Markov chain are intrinsically dependent on prior states. $\tau$ represents specific number of MCMC iterations that emcee sampler requires to generate independent and uncorrelated posterior distributions. This determines the effective sample size $\mathrm{(N_{eff} = N_{total}/\tau)}$, exactly how many independent samples have been gathered to solve a physical model. In our practical implementation, we rigorously quantify the statistical efficiency and absolute convergence of the MCMC sampling by monitoring the dynamic evolution of the integrated autocorrelation time $(\tau)$ throughout the sampler run \ref{fig:tau_all4}. Figure \ref{fig:tau_all4} shows auto-correlation time as a function of the number of walker steps for chiral-odd helicity amplitudes with chiral-even amplitudes fixed within the prior domain $[-10, 10]$. We follow the automated windowing procedure proposed by Goodman \& Weare (2010) \cite{Goodman:2010dyf} in which the parameter chains are periodically evaluated at regular intervals $(N)$ to generate a set of $(\tau)$ values. This allows us to dynamically track and verify whether our sampler has sufficiently explored the multi-dimensional parameter space without being biased from any state. The convergence was monitored by benchmarking the estimated $(\tau)$ values against conservative threshold criteria of $\mathrm{(N \ge 100 \tau )}$. A sTable horizontal plateau obtained below this $\mathrm{(N/100)}$ threshold assures that the total chain length is at least two orders of magnitude larger than the correlation length of successive walker steps, ensuring that the sampler has now gathered an ample effective sample size.

\begin{figure}
    \centering
    \includegraphics[width=0.45\linewidth]{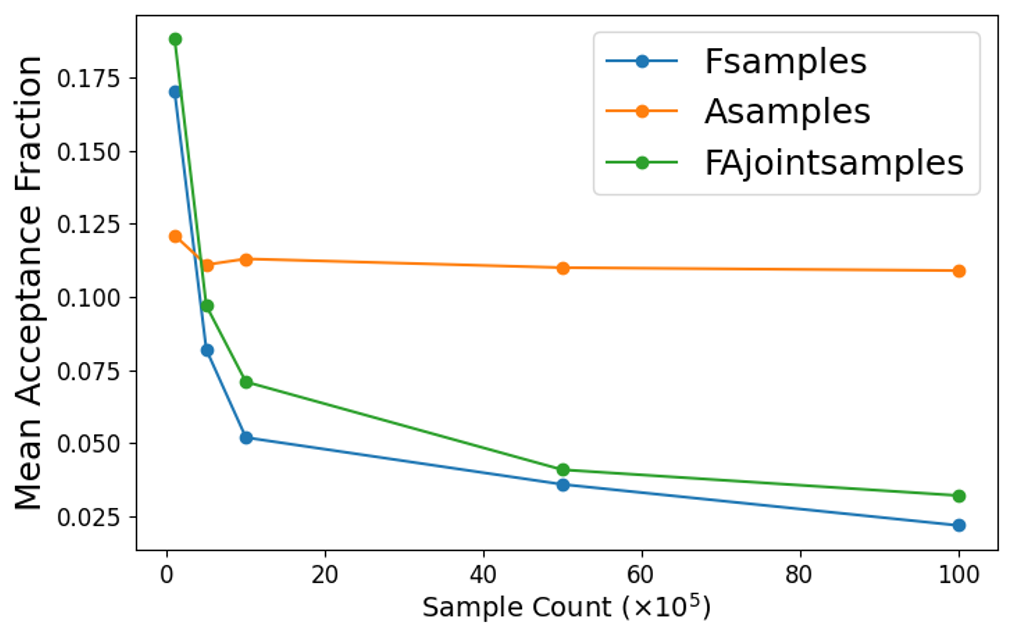}
    \caption{Mean acceptance fraction (MAF) as a function of total sample count $(\times 10^{5})$ for the individual and joint likelihood extraction frameworks where each point is generated from an independent MCMC run at that sample size. The curves compare the independent sampling runs for the chiral-odd helicity amplitudes alone $\mathrm{(\textit{Fsamples}, blue)}$, the spin asymmetries alone $\mathrm{(\textit{Asamples}, orange)}$, and the combined joint likelihood analysis $\mathrm{(\textit{FAjointsamples}, green)}$, illustrating their respective convergence behaviors. chiral-even amplitudes are bounded within the prior domain of $\mathrm{[-10, 10]}$. Correlations correspond to the helicity amplitude MCMC-chain run used to produce samples generated from the combined joint likelihood analysis in shown in Figure \protect\ref{fig:only_HALLB_joint}.}
    \label{fig:maf_all4}
\end{figure}

The mean acceptance fraction (MAF) in emcee is a crucial diagnostic metric which represents the average proportion of proposed MCMC steps accepted by the ensemble of walkers over the course of a sampler run \cite{Goodman:2010dyf}. It is used to evaluate the proposal efficiency and geometric capability of the emcee sampler within chosen set of prior domain. If the MAF is too high, then the walkers are accepting almost every step which implies that the sampler's step size is way too small and the emcee starts behaving like a Gibb's sampler. The walkers in this case take tiny steps taking huge computation time, allowing incredibly slow exploration of the parameter space and montonically increasing $\tau$. If the MAF is too small, the walkers are rejecting almost every step (step size is high). This means that the proposed steps are completely outside the high-probability region. The walkers then stall, wasting computational time and memory being stuck in the same spot which again spikes the $\tau$. For an efficient working emcee, the mean acceptance fraction of the sampler is typically between 0.2-0.5. In our bayesian likelihhod analysis we check this on three scenarios: the likelihood analysis is performed on the cross-section data only, asymmetry data only, and on the joint dataset of cross-section and asymmtery \ref{fig:maf_all4}. In Figure \ref{fig:maf_all4}, we present mean acceptance fraction versus sample-count. Steady MAF for likelihood analysis on asymmetry data only is $(\sim 11\%)$ which is much greater than the other two scenarios. The acceptance rate for Fsamples and FAjoint samples decays rapidly with sample-count \cite{roberts1997weak}. The likelihood analysis on cross-section data only has the least acceptance rate $\sim 2.2\%$ pruning it out from the choice of scenarios to perform a likelihood analysis. The MAF for FAjointsamples is $\sim 3.2\%$. We choose this scenario for our study due to three reasons: (1) The joint analysis not only provides almost uniform, densely filled blobs \ref{fig:only_HALLB_joint} than highly degenerate 2-D plots which we obtain in the scenario of Asamples (although not shown as a part of the paper), (2) We are succesfully able to make the best use of all the available dataset from HALL B, Jefferson Lab, (3) We successfully obtain a convergent integrated auto-correlation time when performing a joint analysis on both cross-section and asymmtery data. 

\subsection{t-dependent Analysis}
\begin{table}[htbp]
\vspace{-2ex}
\centering
\begin{tabular}{|c|c|c|c|c|c|c|c|c|c|c|c|c|c|}
\hline
$Q^2$ & $x_B$ & $\epsilon$ & $-t$ & $\sigma_0$ & $\sigma_{0,stat}$ & $\sigma_{0,sys}$ & $\sigma_{LT}$ & $\sigma_{LT,stat}$ & $\sigma_{LT,sys}$ & $\sigma_{TT}$ & $\sigma_{TT,stat}$ & $s\sigma_{TT,sys}$ \\
\hline
1.87 & 0.270 & 0.598  & 0.12 & 342.0 & $74.0$ & $108.0$ & 1.0 & $86.0$ & $72.0$ & -150.0 & $103.0$ & $101.0$ \\
\hline
1.87 & 0.271 & 0.598  & 0.18 & 437.0 & $54.0$ & $90.0$ & 7.0 & $64.0$ & $74.0$ & 16.0 & $91.0$ & $167.0$ \\
\hline
1.87 & 0.271 & 0.598  & 0.25 & 412.0 & $19.0$ & $32.0$ & 20.0 & $21.0$ & $20.0$ & -233.0 & $34.0$ & $39.0$ \\
\hline
1.87 & 0.271 & 0.598 & 0.49 & 259.5 & $7.30$ & $13.0$ & 25.1 & $7.20$ & $6.10$ & -167.0 & $19.0$ & $14.0$ \\
\hline
2.75 & 0.423 & 0.628 & 0.50 & 323.0 & $19.0$ & $21.0$ & -8.00 & $23.0$ & $16.0$ & -60.0 & $40.0$ & $16.0$ \\
\hline
2.75 & 0.424 & 0.628 & 1.23 & 140.7 & $4.90$ & $9.00$ & -25.8 & $5.60$ & $5.8$ & -16.0 & $13.0$ & $12.0$ \\
\hline
\end{tabular}
\caption{Table of Kinematic bins and cross-section values along with their uncertainties from Jefferson Lab HALL B experiment used for t-dependent Bayesian likelihood analysis \cite{CLAS:2012cna, CLAS:2014jpc}. The notations \text{\it(stat, and sys)} are the statistical and systematic uncertainties, respectively. The kinematic variables $\mathrm{Q^2, (-t)}$ are given in $(\mathrm{GeV^2})$ and the cross-sections are in units of $(\mathrm{nb/GeV^2})$.}
\label{tab:all_sigma_HALLB_data}
\end{table}

Having established and validated the Bayesian likelihood analysis procedure on a single kinematic bin, we extend our analysis and the extraction methodology to a full array of kinematic bins mentioned in Tables \ref{tab:all_sigma_HALLB_data} and \ref{tab:all_asym_data} to investigate the t-dependent behavior of the observables. Table \ref{tab:all_sigma_HALLB_data} presents the list of cross-section data from Jefferson Lab HALL B to be used to study the explicit t-dependence. The kinematic bins very closely match with kinematic bins of asymmetery data from HALL B provided in Table \ref{tab:all_asym_data}. These sets of data allows us to have set of six kinematic bins to be used for the t-dependent analysis. To perform a joint analysis, the bins are selected explicitly based on their matching $(-t)$ values while simultaneously ensuring that the remaining kinematic variables $\mathrm{Q^2, x_{Bj}, \epsilon, W>2}$ remain tightly clustered with negligible variations. The results on each kinematic bin comprising of corner plot on the extracted chiral-odd helicity amplitudes, a consistency check on the model comparing with experimental data, and a sanity check on the helicity amplitudes to calculate the Compton form factors is presented in the supplementary material attached with this manuscript. 

\begin{table}[htbp]
\centering
\begin{tabular}{|c|c|c|c|c|c|c|c|c|c|c|}
\hline
$Q^2$ & $x_B$ & $\epsilon$ & $-t$ & $A_{UL}^{\sin\phi}$ & $A_{UL}^{\sin\phi}\!_{stat}$ & $A_{UL}^{\sin\phi}\!_{sys}$ & $A_{UL}^{\sin 2\phi}$ & $A_{UL}^{\sin 2\phi}\!_{stat}$ & $A_{UL}^{\sin 2\phi}\!_{sys}$ \\
\hline
1.94 & 0.25 & 0.51 & 0.15 & 0.272 & 0.037 & 0.030 & -0.033 & 0.036 & 0.044 \\
\hline
1.94 & 0.25 & 0.51 & 0.26 & 0.223 & 0.033 & 0.060 & -0.047 & 0.032 & 0.030 \\
\hline
1.94 & 0.25 & 0.51 & 0.49 & 0.292 & 0.036 & 0.021 & -0.073 & 0.036 & 0.031 \\
\hline
2.83 & 0.40 & 0.58 & 0.52 & 0.208 & 0.030 & 0.022 & -0.021 & 0.025 & 0.021 \\
\hline
2.83 & 0.40 & 0.58 & 1.24 & 0.164 & 0.049 & 0.051 & -0.090 & 0.042 & 0.036 \\
\hline
\end{tabular}
\vspace{3ex}
\begin{tabular}{|c|c|c|c|c|c|c|c|c|c|c|}
\hline
$Q^2$ & $x_B$ & $\epsilon$ & $-t$ & $A_{LL}^{const}$ & $A_{LL}^{const}\!_{stat}$ & $A_{LL}^{const}\!_{sys}$ & $A_{LL}^{\cos\phi}$ & $A_{LL}^{\cos \phi}\!_{stat}$ & $A_{LL}^{\cos\phi}\!_{sys}$ \\
\hline
1.94 & 0.25 & 0.51 & 0.15 & 0.594 & 0.029 & 0.019 & 0.163 & 0.041 & 0.064 \\
\hline
1.94 & 0.25 & 0.51 & 0.26 & 0.593 & 0.028 & 0.030 & 0.162 & 0.042 & 0.081 \\
\hline
1.94 & 0.25 & 0.51 & 0.49 & 0.564 & 0.036 & 0.031 & 0.200 & 0.064 & 0.078 \\
\hline
2.83 & 0.40 & 0.58 & 0.52 & 0.667 & 0.022 & 0.032 & 0.156 & 0.029 & 0.089 \\
\hline
2.83 & 0.40 & 0.58 & 1.24 & 0.504 & 0.038 & 0.017 & 0.158 & 0.052 & 0.089 \\
\hline
\end{tabular}
\caption{Table of Kinematic bins and Spin Asymmetry $\mathrm{A_{UL} \text{and} A_{LL}}$ values along with their uncertainties from Jefferson Lab HALL B experiment used for for t-dependent Bayesian likelihood analysis \cite{Kim:2015pkf, CLAS:2023wda}. The notations \text{\it(stat, and sys)} are the statistical and systematic uncertainties, respectively. The kinematic variables $\mathrm{Q^2, (-t)}$ are given in $(\mathrm{GeV^2})$.}
\label{tab:all_asym_data}
\end{table}

\begin{figure}
    \centering
    \includegraphics[width=0.75\linewidth]{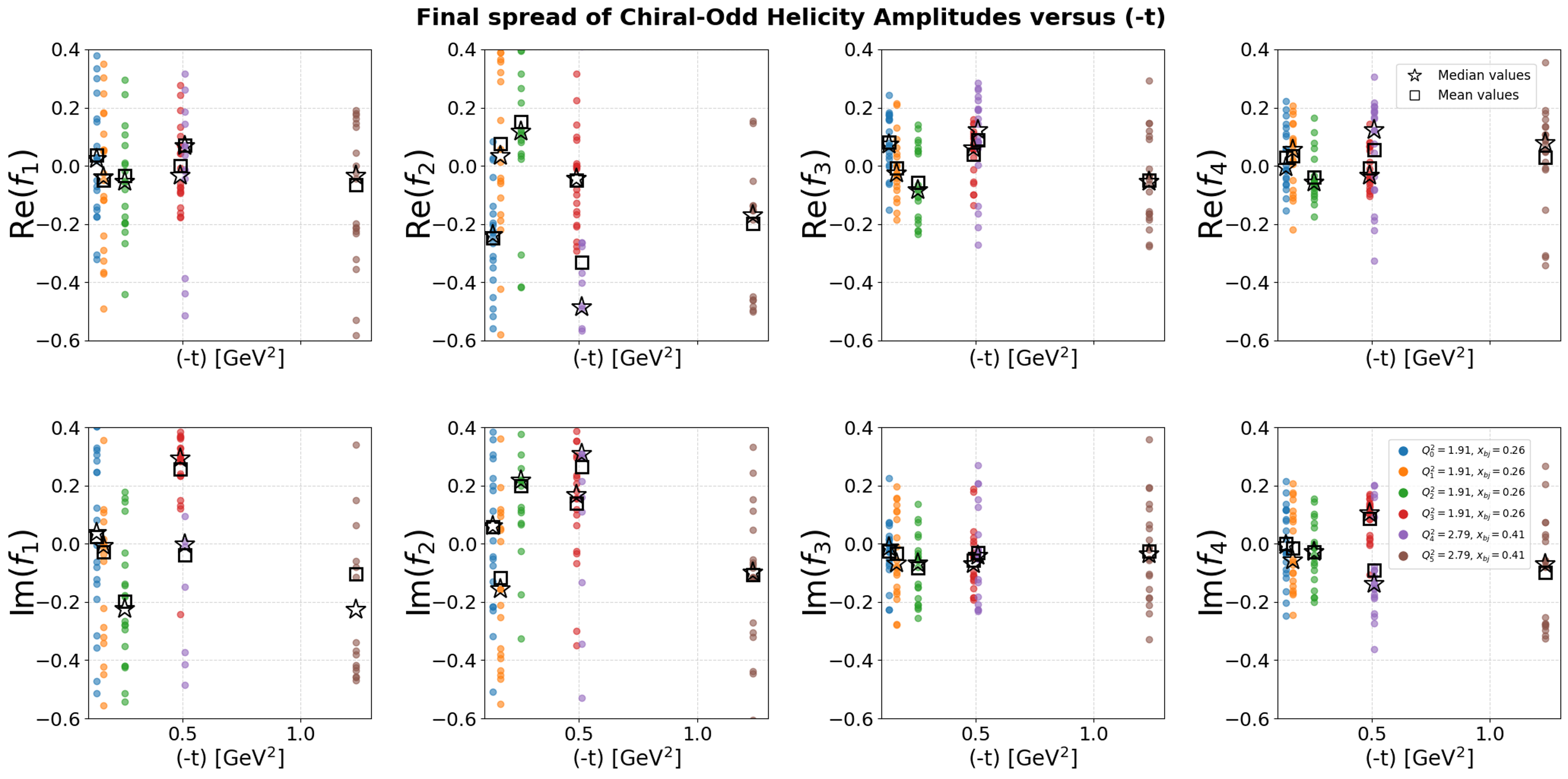}
    \caption{t-dependent extraction of the final 20 chiral-odd helicity amplitudes from the joint likelihood analysis of only Hall B data. The accompanying the chiral even amplitudes are bounded in the prior domain [-10, 10]. The mean and median values at each $\mathrm{(-t)}$ are calculated using the final 20 MCMC chiral-odd helicity amplitude samples generated in Figure \protect\ref{fig:only_HALLB_joint}. Others are also generated in a similar manner.}
    \label{fig:HAs_t_all4}
\end{figure}

In Figure \ref{fig:HAs_t_all4}, we present our extracted chiral-odd helicity amplitudes as a function of (-t). Each panel in the figure illustrates the stability of the extraction, showing the final spread of $20$ post-convergence MCMC samples generated by the joint likelihood analysis.

\begin{figure}
    \centering
    \includegraphics[width=0.75\linewidth]{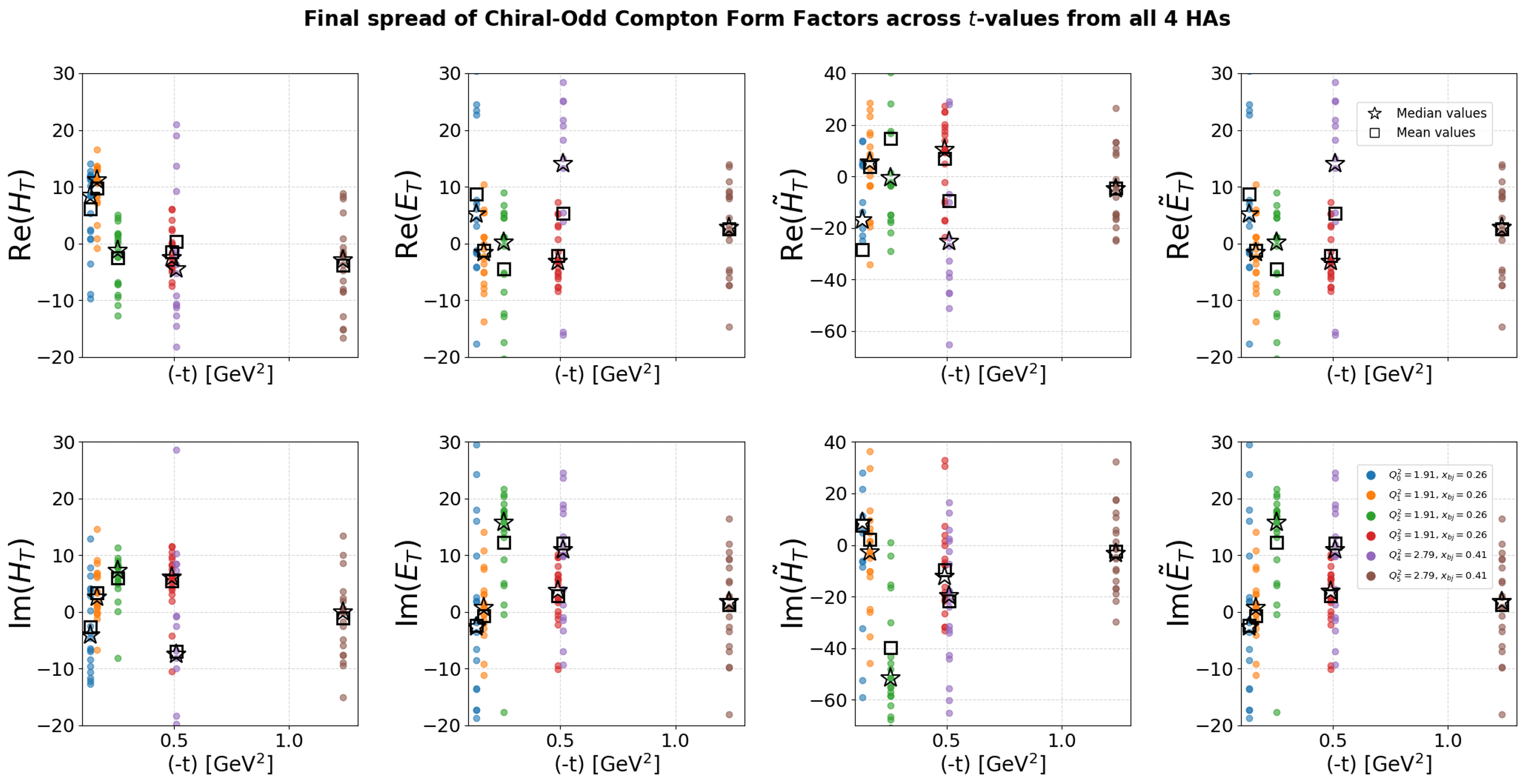}
    \caption{t-dependent extraction of the final 20 chiral Odd Compton-Form Factors generated from chiral-odd helicity amplitude samples from the joint likelihood analysis of only Hall B data. The accompanying the chiral even amplitudes are bounded in the prior domain [-10, 10]. The mean and median values at each $\mathrm{(-t)}$ are calculated using the final 20 MCMC chiral-odd helicity amplitude samples generated in Figure \protect\ref{fig:only_HALLB_joint}. Others are also generated in a similar manner.}
    \label{fig:CFFs_t_all4}
\end{figure}

Following the more statistically correct procedure as discussed above, we extract the final spread of of $20$ post-convergence chiral-odd Compton form factors generated from the sanity checks on the chiral-odd helicity amplitudes obtained from joint likelihood analysis at the six kinematic bins, presented in Figure \ref{fig:CFFs_t_all4}. 

\begin{figure}
    \centering
    \includegraphics[width=0.75\linewidth]{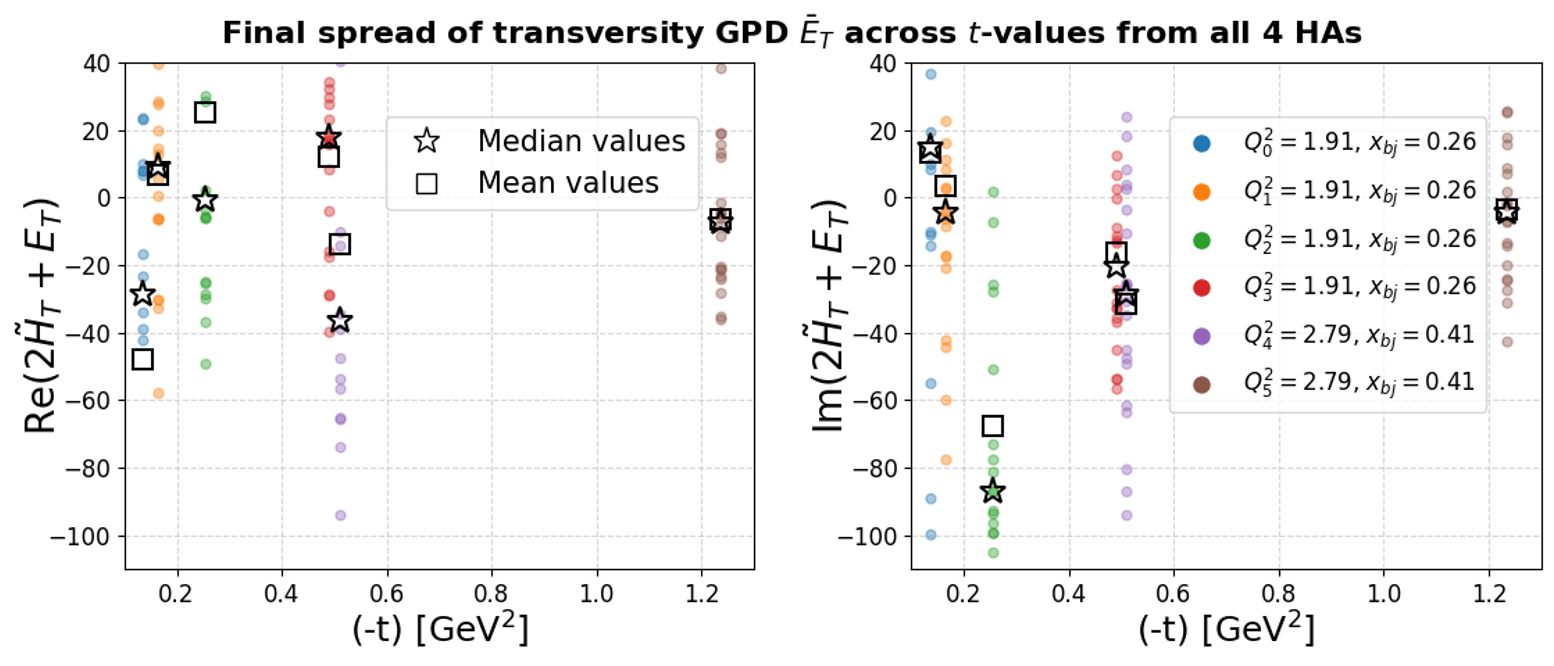}
    \caption{t-dependent extraction of Transversity GPD $\mathrm{\Bar{E}_T = (2\tilde{H}_T + E_T)}$ generated from final 20 chiral Odd Compton-Form Factors which were produced from chiral-odd helicity amplitude samples from the joint likelihood analysis of only Hall B data. The accompanying the chiral even amplitudes are bounded in the prior domain [-10, 10]. The mean and median values at each $\mathrm{(-t)}$ are calculated using the final 20 MCMC chiral-odd helicity amplitude samples generated in Figure \protect\ref{fig:only_HALLB_joint}. Others are also generated in a similar manner.}
    \label{fig:AMM_t_all4}
\end{figure}

In a similar manner, we compute the transversity GPD $\mathrm{\Bar{E}_T = (2\tilde{H}_T + E_T)}$ using the chiral-odd Compton form factors obtained in Figure \ref{fig:CFFs_t_all4}. The real and imaginary parts of $\mathrm{\Bar{E}_T}$ across (-t) values are shown in figure \ref{fig:AMM_t_all4}. To illustrate the qualitative behavior of the observables extracted from the Bayesian likelihood analysis, the median values computed from the final 20 MCMC samples are plotted across all $(-t)$ values. These results are explicitly shown for the chiral-odd helicity amplitudes, Compton form factors (CFFs), and the transversity GPD combination $\mathrm{\Bar{E}_T}$ in Figures \ref{fig:HAs_t_all4}, \ref{fig:CFFs_t_all4}, \ref{fig:AMM_t_all4}, respectively.

\subsubsection{Eliminating one helicity amplitudes}

\begin{figure}[htbp]
    \centering
    \includegraphics[width=0.5\linewidth]{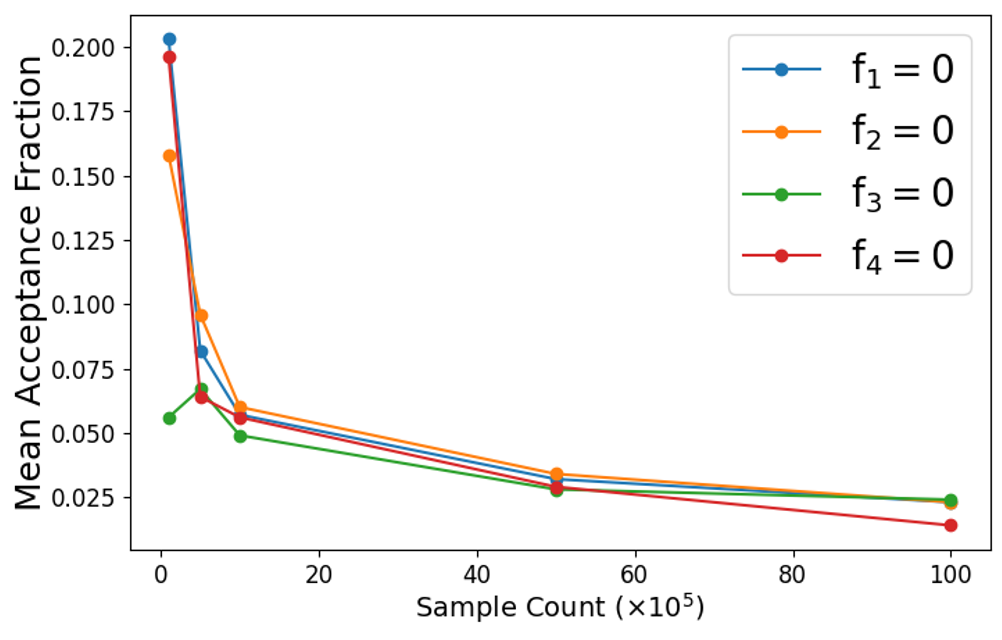}
    \caption{Mean acceptance fraction (MAF) as a function of total sample count $(\times 10^{5})$ for alternative amplitude-truncation configurations. Each point is generated from an independent MCMC run at that sample size. The curves compare the sampling runs for the chiral-odd helicity amplitudes from the combined joint likelihood analysis, illustrating their respective convergence behaviors where each configuration systematically eliminates a single amplitude channel. chiral-even amplitudes are bounded within the prior domain of $\mathrm{[-10, 10]}$.}
    \label{fig:maf_nof2}
\end{figure}

All results shown so far were obtained through an initial joint likelihood analysis incorporating all four chiral-odd helicity amplitudes. However, it still contained persistent multimodality and mild covariance among the posterior samples. Besides this the sanity check performed to extract the chiral-odd Compton form factors were strongly correlated (strong covariance) and highly multi-modal. To alleviate these degeneracies, multi-modal nature, and cross-talk between the parameters, we investigated the effect of truncating the parameter space by eliminating one of the chiral-odd helicity amplitudes. The optimal choice of which amplitude should be eliminated was determined on statistical grounds and systematically evaluating the the mean acceptance fraction plots across all four possible configurations. At the same time, it was ensured that the integrated auto-correlation time remains statistically consistent from the point of convergence at all values of (-t). This will help improving the sampler's mixing and convergence behavior. 

\begin{figure}[htbp]
    \centering
    \includegraphics[width=0.5\linewidth]{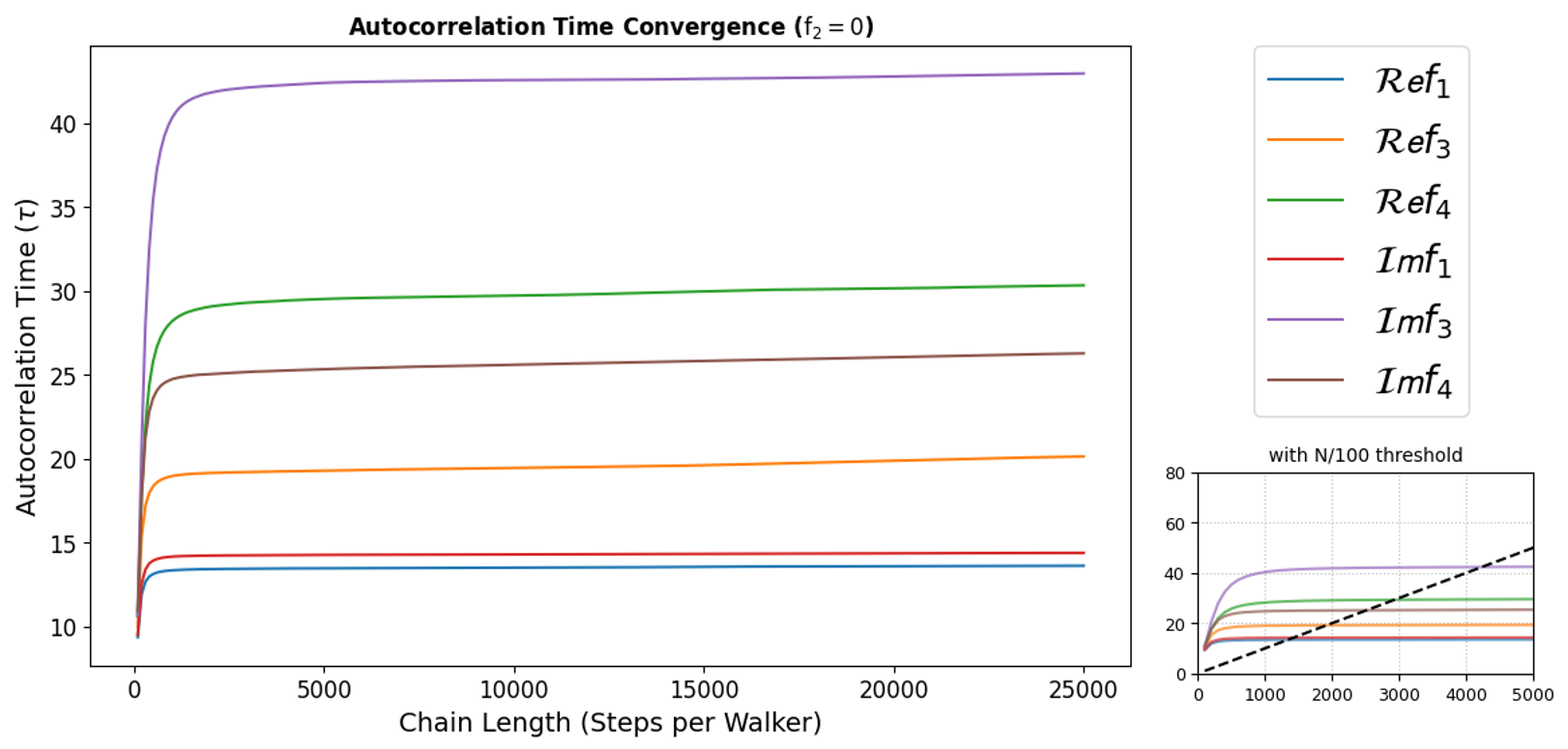}
    \caption{Convergence of the integrated auto-correlation time $\mathrm{\tau}$ as a function of MCMC chain length (steps per walker) for chiral-odd helicity amplitudes under the truncated amplitude configuration $\mathrm{f_2 = 0}$, with chiral-even amplitudes bounded in the prior domain $[-10, 10]$. The main plot illustrates the long-term stabilization of $\mathrm{\tau}$ across walker steps for the various components of the helicity amplitudes during a single MCMC run. The bottom-right inset highlights the early-stage sampling behavior alongside the standard $\mathrm{(N/100)}$ convergence threshold (dashed line). Correlations correspond to the helicity amplitude MCMC-chain run used to produce samples generated in Figure \protect\ref{fig:only_HALLB_joint}.}
    \label{fig:tau_nof2}
\end{figure}

Figure \ref{fig:maf_nof2} shows mean acceptance fraction as a function of sample-count for all 4 possible configurations. In case of $\mathrm{f_1 = 0, f_2 = 0, f_3 = 0}$, the mean acceptance fraction is $\mathrm{\sim 0.02, 0.03, 0.02}$, respectively. $\mathrm{f_1 = 0, f_2 = 0}$ give sTable plateaus while $\mathrm{f_3 = 0}$ has a delayed plateau behavior while $\mathrm{f_4 = 0}$ fails to produce a plateau behavior and acceptance rate cannot be determined. Based on this, the two configurations $\mathrm{f_1 = 0, f_2 = 0}$ are equally likely to be chosen for the analysis. However, the integrated auto-correlation time in these two configurations for all t-values should also be optimal. But in case of $\mathrm{f_1 = 0}$, at some t-values, some of the amplitudes showed spiky behavior. In case of $\mathrm{f_2 = 0}$, at all t-values we obtain sTable plateau behavior as a function of chain length (Figure \ref{fig:tau_nof2}). The threshold line in the inset plot ensures that the convergence is obtained and the sampler has successfully gathered an ample effective sample size. Hence, we use $\mathrm{f_2 = 0}$ configuration to improve our analysis.

\begin{figure}
    \centering
    \includegraphics[width=0.85\linewidth]{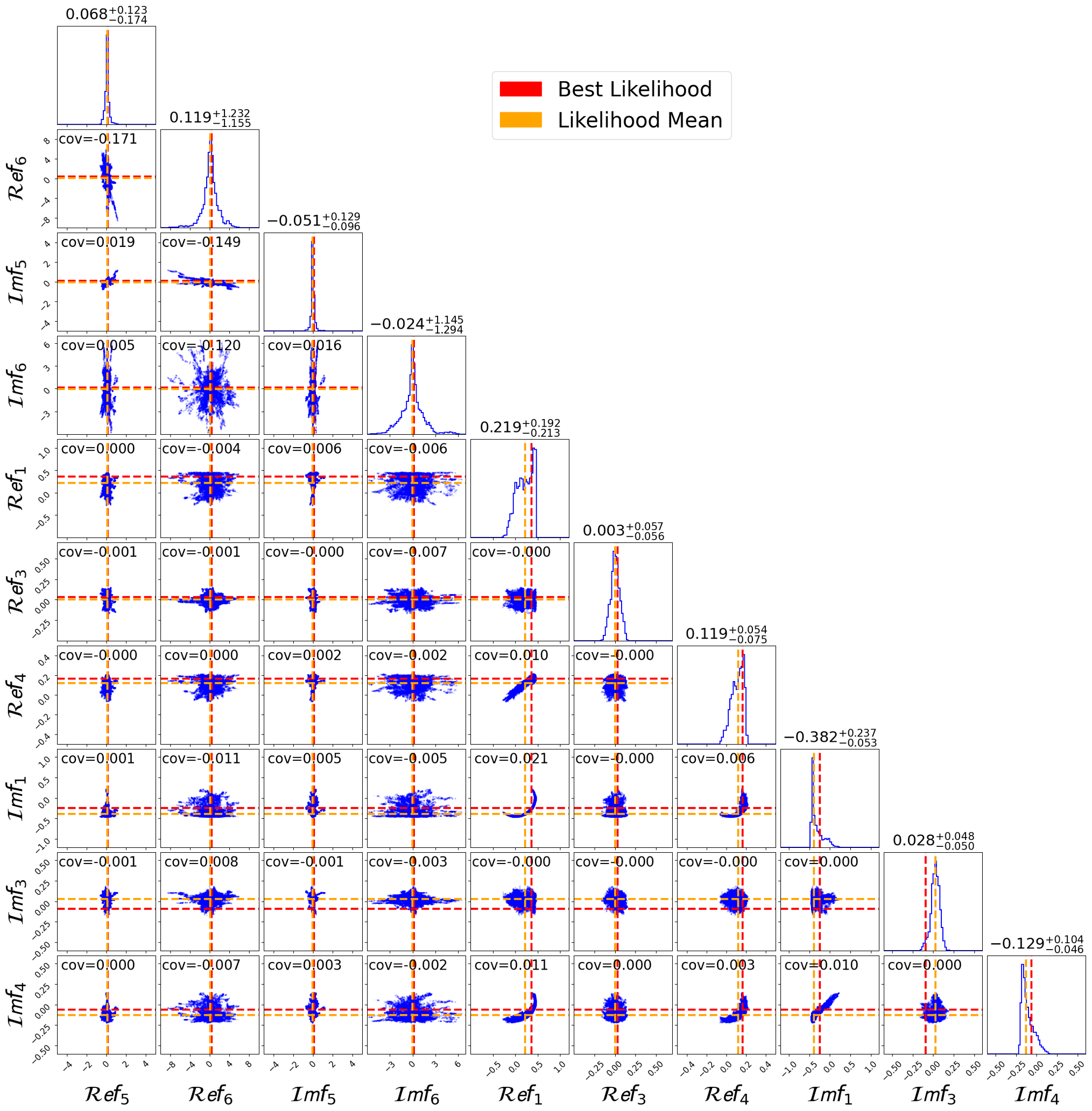}
    \caption{Extraction results for the chiral-odd helicity amplitudes from the joint likelihood analysis of only Hall B data without Beam Spin Asymmetry $\mathrm{A_{LU}}$ under the truncated amplitude configuration $\mathrm{f_1, f_3, f_4 \neq 0}$ and $\mathrm{f_2 = 0}$. The accompanying chiral-even amplitudes are bounded within a uniform prior domain of $\mathrm{[-10,10]}$.}
    \label{fig:t3_nof2_HAs}
\end{figure}

\begin{figure}
    \centering
    \includegraphics[width=0.85\linewidth]{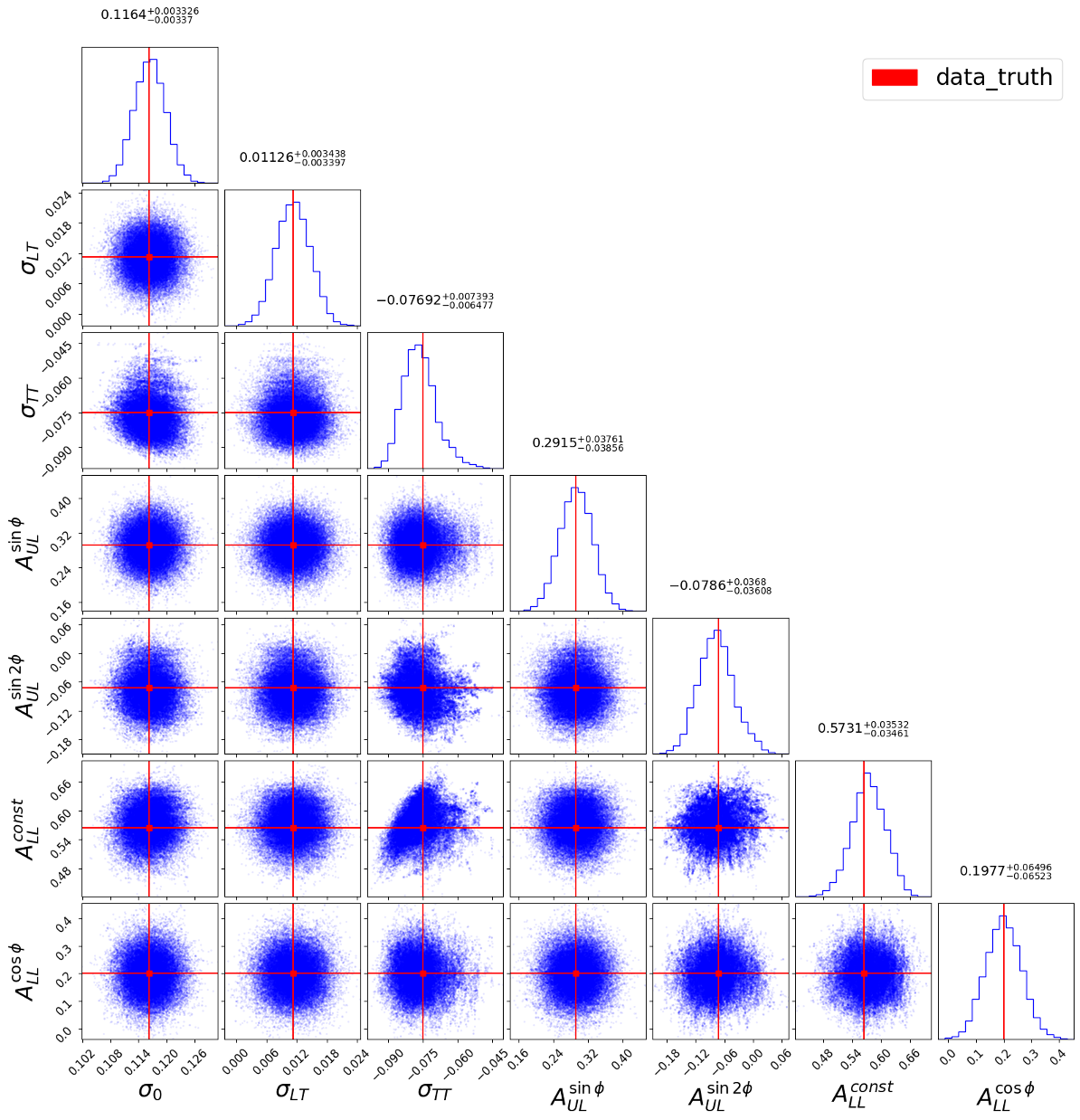}
    \caption{Consistency check on the experimental observables with the extracted chiral-odd helicity amplitudes using joint likelihood analysis of alone Hall B data without Beam Spin Asymmetry $\mathrm{A_{LU}}$ presented in Figure \ref{fig:t3_nof2_HAs}. The accompanying chiral-even amplitudes are constrained within a uniform prior domain of $\mathrm{[-10,10]}$.}
    \label{fig:cc_t3_nof2_HAs}
\end{figure}

Figure \ref{fig:t3_nof2_HAs} shows extraction of chiral-odd helicity amplitudes for $\mathrm{f_2 = 0}$ configuration along with the consistency check on the MCMC samples obtained from the joint analysis of cross-section and asymmetry data from HALL B. The kinematic here chosen is same as that used for case with all four amplitudes \ref{fig:only_HALLB_joint}. The consistency check corner plot on the right panel of the figure shows strong agreement with experimental data showing a good predictive power of the model and the emcee algorithm. The 1-D histograms are unimodal compared to the case with all 4 amplitudes \ref{fig:only_HALLB_joint} and 2-D plots also show no covariance comparitively. The CFFs obtained (please see the supplementary material) less correlated (huge fall in covariance) and the 1-D histograms are also unimodal. Thus, $\mathrm{f_2 = 0}$ configuration indeed helps to solve the statistical discrepancies which were arising in the case of all four amplitudes.

\begin{figure}
    \centering
    \includegraphics[width=0.75\linewidth]{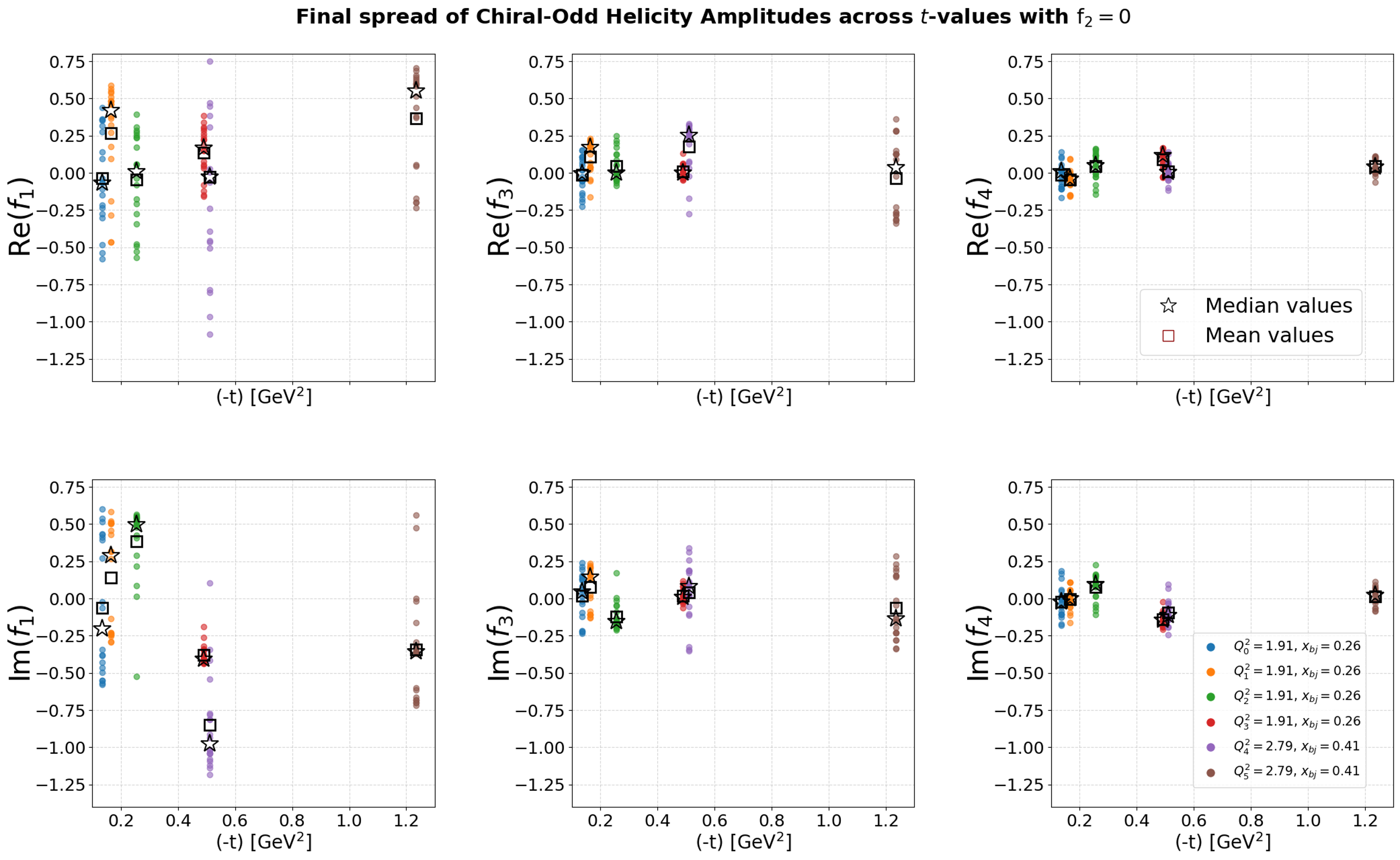}
    \caption{t-dependent extraction of chiral-odd helicity amplitudes from the joint likelihood analysis of only Hall B data under the truncated amplitude configuration $\mathrm{f_2 = 0}$. The accompanying the chiral even amplitudes are bounded in the prior domain [-10, 10]. The mean and median values at each $\mathrm{(-t)}$ are calculated using the final 20 MCMC chiral-odd helicity amplitude samples generated in Figure \protect\ref{fig:only_HALLB_joint}. Others are also generated in a similar manner.}
    \label{fig:nof2_HAs_tvals}
\end{figure}

\begin{figure}
    \centering
    \includegraphics[width=0.75\linewidth]{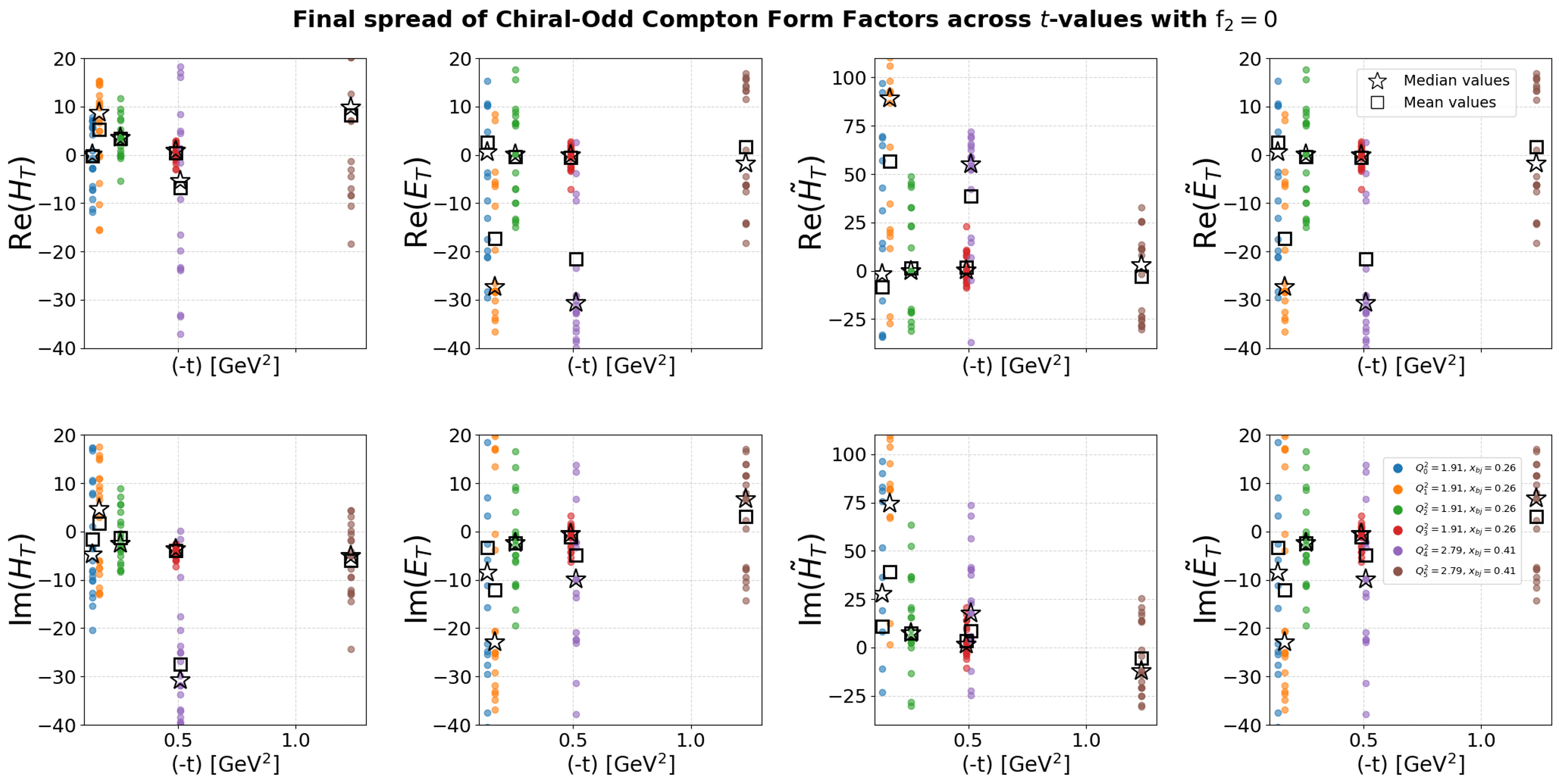}
    \caption{t-dependent extraction of chiral Odd Compton-Form Factors generated from chiral-odd helicity amplitude samples from the joint likelihood analysis of only Hall B data under the truncated amplitude configuration $\mathrm{f_2 = 0}$. The accompanying the chiral even amplitudes are bounded in the prior domain [-10, 10]. The mean and median values at each $\mathrm{(-t)}$ are calculated using the final 20 MCMC chiral-odd helicity amplitude samples generated in Figure \protect\ref{fig:only_HALLB_joint}. Others are also generated in a similar manner.}
    \label{fig:nof2_CFFs_tvals}
\end{figure}

A t-dependent analysis was again performed for the $\mathrm{f_2 = 0}$ configuration to obtain more robust solutions. Figures \ref{fig:nof2_HAs_tvals} and \ref{fig:nof2_CFFs_tvals} present the final spread of 20 post-convergence MCMC samples generated by the joint Bayesian likelihood analysis. The sampler was run on the helicity amplitudes (forward extraction as before to be statistically motivated) and the chiral-odd Compton form factors were obtained as a sanity check on the analysis.

\begin{figure}
    \centering
    \includegraphics[width=0.75\linewidth]{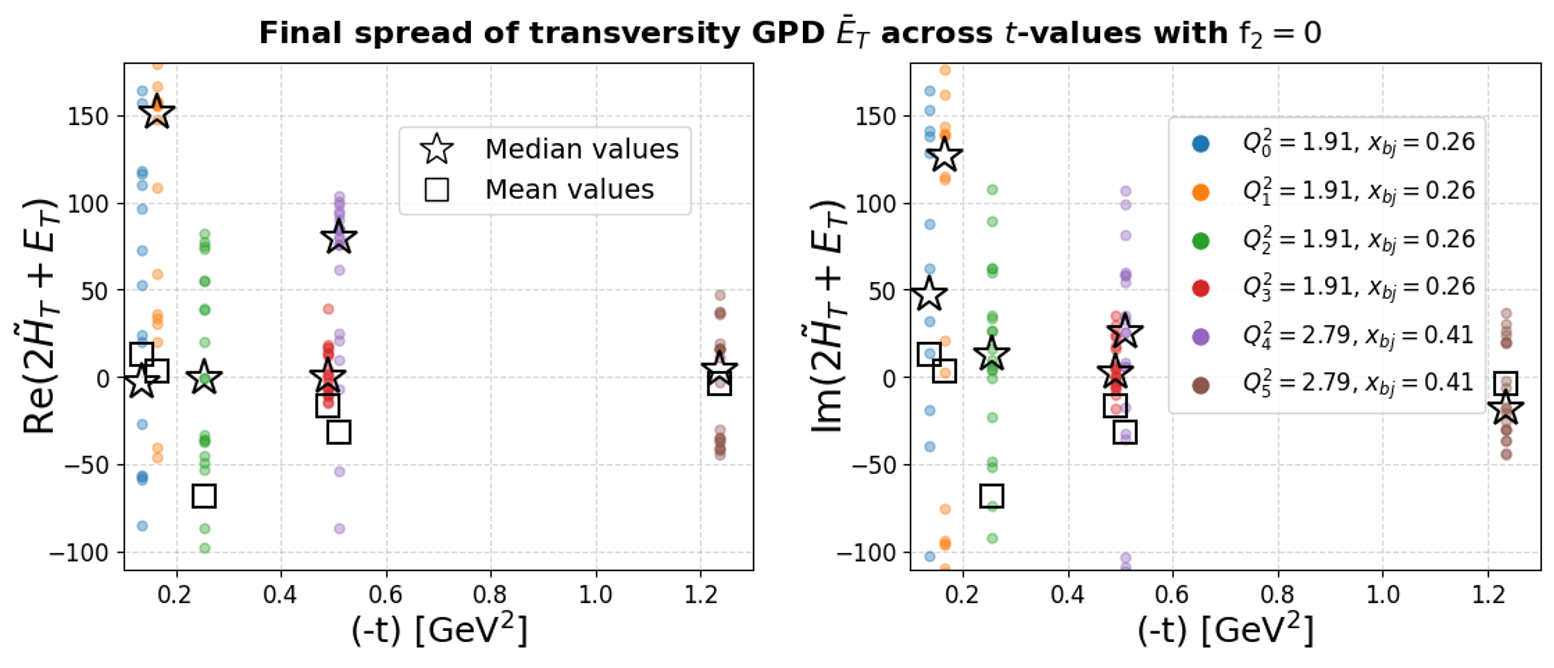}
    \caption{t-dependent extraction of Transversity GPD $\mathrm{\Bar{E}_T = (2\tilde{H}_T + E_T)}$ generated from chiral Odd Compton-Form Factors which were produced from chiral-odd helicity amplitude samples from the joint likelihood analysis of only Hall B data  under the truncated amplitude configuration $\mathrm{f_2 = 0}$. The accompanying the chiral even amplitudes are bounded in the prior domain [-10, 10]. The mean and median values at each $\mathrm{(-t)}$ are calculated using the last 20 MCMC chiral-odd helicity amplitude samples generated in Figure \protect\ref{fig:only_HALLB_joint}. Others are also generated in a similar manner.}
    \label{fig:nof2_AMM_t_all4}
\end{figure}

In a similar manner, the transversity GPD $\mathrm{\Bar{E}_T = (2\tilde{H}_T + E_T)}$ was also obtained from the MCMC generated chiral-odd CFFs (see Figure \ref{fig:nof2_AMM_t_all4}). To understand the qualitative behavior in (-t), median values at each point in (-t) were calculated from final spread of 20 MCMC samples. These extractions will demonstrate the viability of extracting the fundamental tensor charge and tensor anomalous magnetic moment of the proton from real experimental data, establishing a clear pathway for future phenomenological analyses. We see this in detail in the next subsection.

\subsection{Extraction of Tensor Charge and Tensor Anomalous Magnetic Moment}

\begin{figure}
    \centering
    \includegraphics[width=0.65\linewidth]{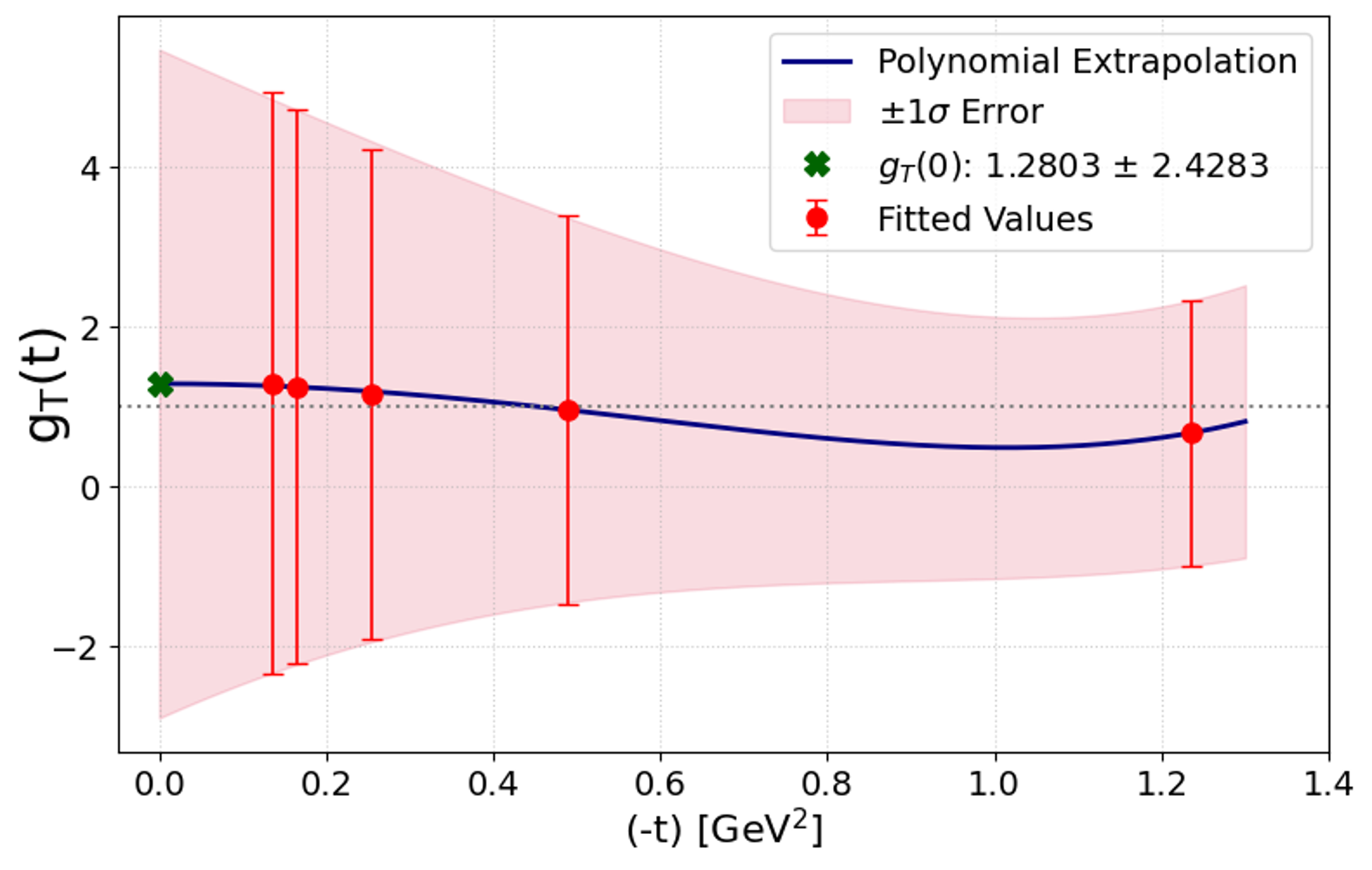}
    \caption{Preliminary extraction of Tensor charge $g_T (t)$ from joint Bayesian likelihood analysis on HALL B data as a function of $\mathrm{(-t)}$. The red points shown are the calculated $\mathrm{g_T (t)}$ values obtained after integration using the mean of final 20 MCMC samples on chiral-odd Compton form factors presented in Figure \ref{fig:CFFs_t_all4} for all t-values. Equation \ref{eq:fi_form} fitted to the mean values of final 20 MCMC samples from Figure \ref{fig:CFFs_t_all4}, and integrated to calculate $\mathrm{g_T (t)}$, and error is propagated by assuming gaussian errors in Figure \ref{fig:CFFs_t_all4}.}
    \label{fig:gt_tvalues}
\end{figure}

In this subsection, we present a possible extraction of the proton's tensor charge and tensor anomalous magnetic moment from the physics observables we extracted using the joint Bayesian likelihood analysis on experimental data from Jefferson Lab, HALL B. We use a Reggeized parameteric form to study the t-dependence of the extracted chiral-odd Compton form factors which when extrapolated give the fundamental tensor charge ($\mathrm{g_T (0)}$) and the tensor anomalous magnetic moment ($\mathrm{\kappa_T (0)}$). The parametric form used is as follows:
\begin{equation}
    H_T(X, \xi, t) = \mathcal{N}(\xi)\;(X^{-(\alpha + \alpha'(1 - X)^p\;t)}  \quad \Rightarrow \quad g_T = \int_{-1}^1 H_T(x,0,0) \, dX 
    \label{eq:fi_form}
\end{equation}
where, $\mathrm{\mathcal{N}(\xi), \alpha, \alpha', and p}$ are the fit parameters (similarly for ${\Bar{E}_T}$. 
It should be noticed that, assuming that only the $u$ and $d$ quarks contribute to the GPD, what is extracted from $\pi^o$ electroproedution off a proton target is the combination,
\[ g_T = \frac{1}{\sqrt{2}} \left( \frac{2}{3} g_T^u + \frac{1}{3} g_T^d \right) ,\]
and that, therefore, this results is not directly applicable to the BSM search of {\it e.g.} Ref.\cite{Beck:2023hnt}, which is sensitive to $g_T^u - g_T^d$. The availability of additional deeply virtual $\eta$ meson electroproduction data \cite{CLAS:2017jjr} using a proton polarized target would however allow us to perform a flavor separated analysis of the tensor charge \cite{Goldstein:2014aja}.
\begin{table}[htbp]
\centering
\caption{Fitting Parameters obtained from Hessian-based fitting method performed on the mean values calculated using the final 20 MCMC samples on chiral-odd Compton form factors presented in Figure \ref{fig:CFFs_t_all4} for all t-values.}
\label{tab:fit_params}
\begin{tabular}{|c|c|c|c|c|c|c|}
\hline
$\mathcal{N}_g$ & $\alpha_g$ & $\alpha'_g$ & $p_g$ \\
\hline
$0.6509 \pm0.8895$ & $0.5662 \pm1.1369$ & $0.6104 \pm2.0004$ & 1.0  \\
\hline
\end{tabular}
\vspace{1ex}
\\
\begin{tabular}{|c|c|c|c|c|c|c|}
\hline
$\mathcal{N}_\kappa$ & $\alpha_\kappa$ & $\alpha'_\kappa$ & $p_\kappa$ \\
\hline
$20 \pm1.414$ & $-4.0 \pm6.6327$ & $5.0 \pm46.4709$ & 1.0 \\
\hline
\end{tabular}
\end{table}


\begin{figure}
    \centering
    \includegraphics[width=0.65\linewidth]{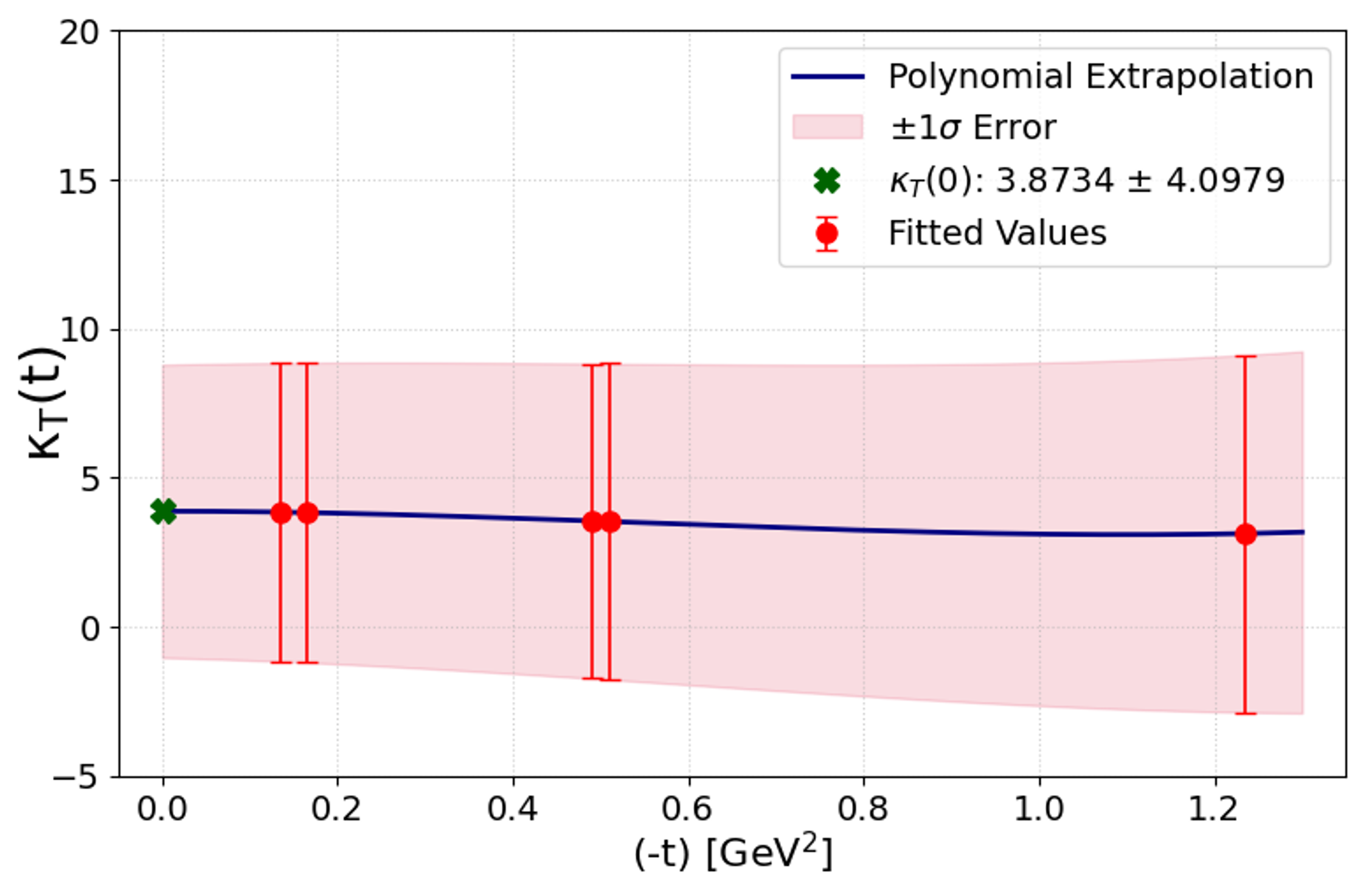}
    \caption{Preliminary extraction of Tensor Anomalous Magnetic Moment ($\mathrm{\kappa_T (t)}$) from joint Bayesian likelihood analysis on HALL B data as a function of $\mathrm{(-t)}$. The red points shown are the calculated $\mathrm{\kappa_T (t)}$ values obtained after integration using the mean of final 20 MCMC samples on chiral-odd Compton form factors presented in Figure \ref{fig:CFFs_t_all4} for all t-values. The errors are propagated by assuming gaussian errors in figure \ref{fig:CFFs_t_all4}.}
    \label{fig:kt_tvalues}
\end{figure}

Using this parametric form we use a Hessian-based optimization method to fit the extracted Transversity GPDs $\mathrm{H_T}$ and $\mathrm{\Bar{E}_T}$ in Figures \ref{fig:CFFs_t_all4} and \ref{fig:AMM_t_all4} to obtain the fitting parameters at $\mathrm{\zeta = 0.26}$ presented in Table \ref{tab:fit_params}. These parameters were used to calculate the $\mathrm{(-t)}$ dependence of the integrated tensor charge ($\mathrm{g_T (t)}$) and the tensor anomalous magnetic moment ($\mathrm{\kappa_T (t)}$) using the same Reggeized model as \ref{eq:fi_form} which were then extrapolated using a polynomial fit to calculate integrated proton's tensor charge ($\mathrm{g_T (0)}$) and the tensor anomalous magnetic moment ($\mathrm{\kappa_T (0)}$). The results obtained are presented in Figures \ref{fig:gt_tvalues} and \ref{fig:kt_tvalues}. These quantities with their uncertainties are mentioned below in Table \ref{tab:gt_kt}. It is interesting to see that the extracted quantities have exceedingly large error bounds. This is attributed to the fact that our joint analysis was performed on a single kinematic bin each time. The other kinematics $\mathrm{Q^2, x_{Bj}, t}$ were close enough to perform a bayesian analysis, but at the same time far enough to impact the final outcomes. This points out an important neccessity of more experimental data at a particular kinematic bin to start with.

\begin{table}[htbp]
\centering
\begin{tabular}{|c|c|}
\hline
$\mathrm{g_T (0)}$ & $\mathrm{\kappa_T (0)}$ \\
\hline
$1.28 \pm2.43$ & $3.87 \pm4.10$ \\
\hline
\end{tabular}
\caption{Tensor charge ($\mathrm{g_T (0)}$) and the tensor anomalous magnetic moment ($\mathrm{\kappa_T (0)}$) obtained from polynomial extrapolation performed on the to compute at $\mathrm{t=0}$ (green star in the Figures \ref{fig:gt_tvalues} and \ref{fig:kt_tvalues}) values calculated using the final 20 MCMC samples on chiral-odd Compton form factors presented in Figure \ref{fig:CFFs_t_all4} for all t-values.}
\label{tab:gt_kt}
\end{table}

\section{Conclusions/Summary}
\label{sec:summary}
In summary, we presented a systematic study of chiral-odd GPDs leading to an extraction of the tensor charge from pseudoscalar $\mathrm{\pi^0}$ production at Jefferson Lab. By implementing a rigorous Bayesian likelihood framework on Jefferson Lab datasets, we have successfully resolved the underlying constraints and structural discrepancies which are otherwise inaccessible through traditional, single-glance analysis techniques such as Hessian-based extractions. We find that the non-linear mathematical structure of the cross-section renders the extraction  of the chiral-odd GPDs from experimental data intrinsically nontrivial. 
\\

{\it Let alone, the model cannot be responsible for extraction with poor statistical precision. }
\\

At the same time, the interpretation of these constraints depends on the theoretical description of the reaction mechanism. In particular, the connection between the measured observables and the chiral-odd GPDs relies on assumptions concerning the dominance of transverse amplitudes and the applicability of the factorized description used for pseudoscalar meson production. Our results therefore provide not only quantitative constraints on the chiral-odd sector, but also a framework for assessing the extent to which the presently available data are consistent with these underlying theoretical assumptions.

\noindent The large uncertainty in the  extraction is inherent to the sparse kinematic coverage and the structural degeneracies of the cross-section. The deficiency in the amount of data and type of experimental data severely affects the extraction of fundamental quantities such as the tensor charge.
This detailed analysis also determines what combination of experimental data works best for the analysis. 
Furthermore, our findings demonstrate that a direct extraction of core physical parameters, chiral-odd helicity amplitudes from the measured observables (experimental data) significantly improves the statistical precision, avoiding the additional model dependence and error propagation inherent in first isolating the chiral-odd CFFs (reverse extraction). 
Concurrently, our analysis yields an intriguing result: reducing the non-linear structure of the cross-section through the exclusion of one of the helicity amplitudes improves the extraction of the fundamental quantities. Our Bayesian likelihood analysis also confirms that the fundamental tensor charge and tensor anomalous magnetic moment can be reliably extracted from real experimental datasets using this joint analysis framework. However, this analysis could become more statistically precise if we include:

\begin{enumerate}
    \item More experimental data on the same observables at the same kinematic bin $\mathrm{[E,Q^2, x_{Bj}, t, \epsilon]}$.
    \item More polarization configurations to help resolving the non-linear behavior of the cross-section model when extracting the helicity amplitudes. In particular, this analysis would highly benefit from having experimental data for a {\it transversely polarized target}, in addition to the longitudinal, to guarantee  a precise extraction of the observables.
    \item Under the premise that each GPD is unique at a given kinematic bin, the joint likelihood analysis could provide a more precise and systematic extraction of the underlying physical quantities if we had the same data on the structure functions and asymmetries with their full $\mathrm{\phi}$ dependence. 
    \item The MCMC samples exhibit a pronounced ``clumpy'' structure, characterized by distinct concentrations of samples in different regions of parameter space rather than a single, smoothly connected posterior region. Such clustering reflects the presence of multiple parameter combinations that are compatible with the available experimental constraints
    The clumpiness in the MCMC samples arises from two main factors: 
    \begin{enumerate}
\item  the limited amount of experimental data. So far we have used one data point on each structure function at a single kinematic bin. The limited data 
does not sufficiently restrict the allowed parameter space, permitting the walkers to populate multiple regions corresponding to different combinations of helicity amplitudes that provide comparably good descriptions of the data.
\item even with less data, the MCMC requires huge computational power. 
For the forward extraction, we estimate that approximately  $10^{12}$ samples would be required to achieve the desired statistical sampling of the posterior distribution, making a direct calculation at this scale computationally expensive. 
%
The reverse extraction is considerably more demanding: within the range tested, up to $10^{8}$ samples, its computational cost is approximately a factor of $18$ larger than that of the forward extraction. Extending the reverse analysis toward sample counts of $10^{12}$--$10^{15}$ would therefore present a substantial computational challenge. 
While computational scaling techniques such as multiprocessing and high-performance computing (HPC) can reduce the absolute execution time, they cannot overcome the fundamental information bottlenecks of the fit unless we have the right amount and type of experimental data: to overcome these issues the Bayesian Likelihood analysis maps out the parameter dependence and observables in the most transparent way. 
\end{enumerate}
\end{enumerate}

Experiments cannot freely choose identical kinematics for every measurement, since the accessible phase space is constrained by the detector acceptance, experimental configuration, and available statistics. Consequently, observables that are individually sensitive to the same underlying quantities may not be directly accessed in a model-independent, bin-by-bin extraction. A reasonable goal for the interplay between experiment and phenomenology is therefore to identify, within experimentally accessible kinematics, sets of complementary observables that can be measured at sufficiently overlapping kinematic points to maximize their combined constraining power. The present analysis constitutes a step towards this goal.


\begin{acknowledgments}
This work was completed under the Department of Energy grant DE-SC0024644 and DOE grant DE-SC0016286 (D.A., S.L., Z.P.). We thank all members of the ExclAIm Collaboration for providing a stimulating environment and for many fruitful discussions. We are also grateful to Whitney Armstrong for valuable discussions on the interplay between experimental measurements and phenomenological analyses, and to Zhoudunming Tu (Kong) and Salvatore Fazio for discussions during our stay at the EIC Theory Group Institute in the summer of 2026.

\end{acknowledgments}

\appendix

\section{Pion vertex form factors}
\label{app_g}
\begin{widetext}
\begin{eqnarray}
g_\pi^{V,odd}  =  \int dx_1 dy_1 \int  d^2 b  
\, \hat{\psi}_V(y_1,b) \, K_o(x_1 \sqrt{1-x_1} Q^2 b) \, \left[ 8 C_F \alpha_S(x_1 y_1 Q^2) \right]
 \exp[-S]     \, \hat{\phi}_{\pi^o}(x_1,b)  && \\
g_\pi^{A, odd}=  \int dx_1 dy_1 \int d^2  b  
\, \hat{\psi}_A(y_1,b) \, K_o(x_1 \sqrt{1-x_1} Q^2 b)  \, \left[ 8 C_F \alpha_S(x_1 y_1 Q^2) \right]
\exp[-S]  \, \hat{\phi}_{\pi^o}(x_1,b) &&
\end{eqnarray}
\end{widetext}
where, 
\begin{equation}
\hat{\psi}_{A}(y_1,b) = \int d^2 k_T J_1(y_1 b) \psi_V(y_1,k_T) \, .
\end{equation}
$x_1(y_1)$ is the longitudinal momentum fraction carried by the quark, $b$ is the Fourier transform of the transverse intrinsic momentum, $k_T$,  $\psi_{V}(y_1,b)$ and $\psi_{\pi^o}(x_1,b)$ are the vector meson and pion wave functions in configuration space, respectively, $K_o$, the modified Bessel function of order 
zero, results from the Fourier transform of the hard scattering amplitude,  and
$exp(-S)$ is the Sudakov exponential. 
It is important to observe that it is sufficient for our purpose to use the leading twist pion wave function since the power suppression 
due to the $\gamma^5$ coupling is already accounted for by the photon polarization vector. 

Furthermore, in the vector channel the initial and final states have the same orbital angular momentum, $L=0$, whereas the axial-vector channel involves a transition from $L=1$ in the initial state to $L=0$ in the final state. The presence of nonzero orbital angular momentum in the initial state gives rise to a higher-order Bessel function in the corresponding impact-parameter representation. This structure explicitly accounts for the difference between OAM-conserving ($L=0\rightarrow0$) and OAM-changing ($L=1\rightarrow0$) transitions. Since the higher-order Bessel function originates from the $L=1$ component, it modifies the transverse spatial weighting of the axial-vector contribution and enhances its sensitivity to configurations at larger transverse distances.
The vector ($V$) and axial-vector ($A$) contributions enter the helicity amplitudes according to
\[
f_{10}^{++},,f_{10}^{--}\propto g^V,\qquad
f_{10}^{+-}\propto g^V+g^A,\qquad
f_{10}^{-+}\propto g^V-g^A.
\]

\bibliography{alu_bib}

\end{document}